\documentclass[12pt]{article}
\usepackage[margin=1.in]{geometry}
\usepackage[utf8]{inputenc}
\usepackage{amsthm,amsmath,physics,mathtools,amssymb}%mathrsfs,amsfonts,

\usepackage{charter}
\usepackage{CJKutf8}

\usepackage{enumitem}

\usepackage{hyperref}
\hypersetup{
    colorlinks=true,
    linkcolor=blue,
    filecolor=magenta,      
    urlcolor=cyan,
}
\usepackage[capitalise]{cleveref}

\theoremstyle{definition}
\usepackage{graphicx}
\graphicspath{{figures/}}

\usepackage{tensor}

\DeclarePairedDelimiterX{\inp}[2]{\langle}{\rangle}{#1, #2}

\usepackage{xparse}

\NewDocumentCommand\LH{mo}{%
  \IfNoValueTF{#2}
   {\mathcal{L}(\mathcal{H}^{#1})}
   {\mathcal{L}(\mathcal{H}^{#1},\mathcal{H}^{#2})}%
}

\newcommand\id{\leavevmode\hbox{\small1\kern-3.3pt\normalsize1}}

\newcommand{\sV}{\mathbb{V}}

\allowdisplaybreaks

\usepackage{bbold}

\usepackage{authblk}

\usepackage[title]{appendix}

\usepackage{mciteplus}

\usepackage{enumitem} 

\usepackage{float}

\title{Light ray fluctuation and lattice refinement of simplicial quantum gravity}

\author{Ding Jia (贾丁)\thanks{djia@perimeterinstitute.ca}}
\affil{Perimeter Institute for Theoretical Physics, Waterloo, Ontario, N2L 2Y5, Canada}
\affil{Department of Physics and Astronomy, University of Waterloo, Waterloo, Ontario, N2L 3G1, Canada}
\date{}

\begin{document}

\begin{CJK*}{UTF8}{gbsn}
\maketitle
\end{CJK*}

\begin{abstract}
In several approaches of non-perturbative quantum gravity, a major outstanding problem is to obtain results valid at the infinite lattice refinement limit. Working with Lorentzian simplicial quantum gravity, we compute light ray fluctuation probabilities in 3D and 4D across different lattices. In a simplified refined box model with the Einstein-Hilbert action, numerical results show that lattice refinement does not simply suppress or simply enhance light ray fluctuations, but actually drives very wide and very narrow light probability distributions towards intermediate ones. A comparison across lattices and across couplings reveals numerical hints at a lattice refinement fixed point associated with a universality class of couplings. The results fit the intuition that quantum spacetime fluctuations reflected by light ray fluctuations start wild microscopically and become mild macroscopically. The refined box model is limited by the assumption of a rigid frame at all scales. The present results suggest further studies around the zero-coupling limit to relax the simplifying assumptions of the model.
\end{abstract}

\section{Introduction}

In quantum gravity, a major challenge is to identify interesting physical quantities to compute on the theory side and to measure on the experiment side. Another major challenge is to make non-perturbatively defined theories work at the practical level. In particular, while theories such as Lorentzian simplicial quantum gravity \cite{Jia2022ComplexProspects}, (locally) causal dynamical triangulation \cite{Ambjorn2012NonperturbativeGravity, Loll2020QuantumReview}, and spinfoam models \cite{Perez2013TheGravity, Rovelli2014CovariantGravity} may appear promising, they face the common outstanding task of taking the lattice refinement limit \cite{Jia2022LightGravity, Ambjorn2020RenormalizationGeometryb, Steinhaus2020CoarseReviewb}.

\begin{figure}%[h]
    \centering
    \includegraphics[width=.4\textwidth]{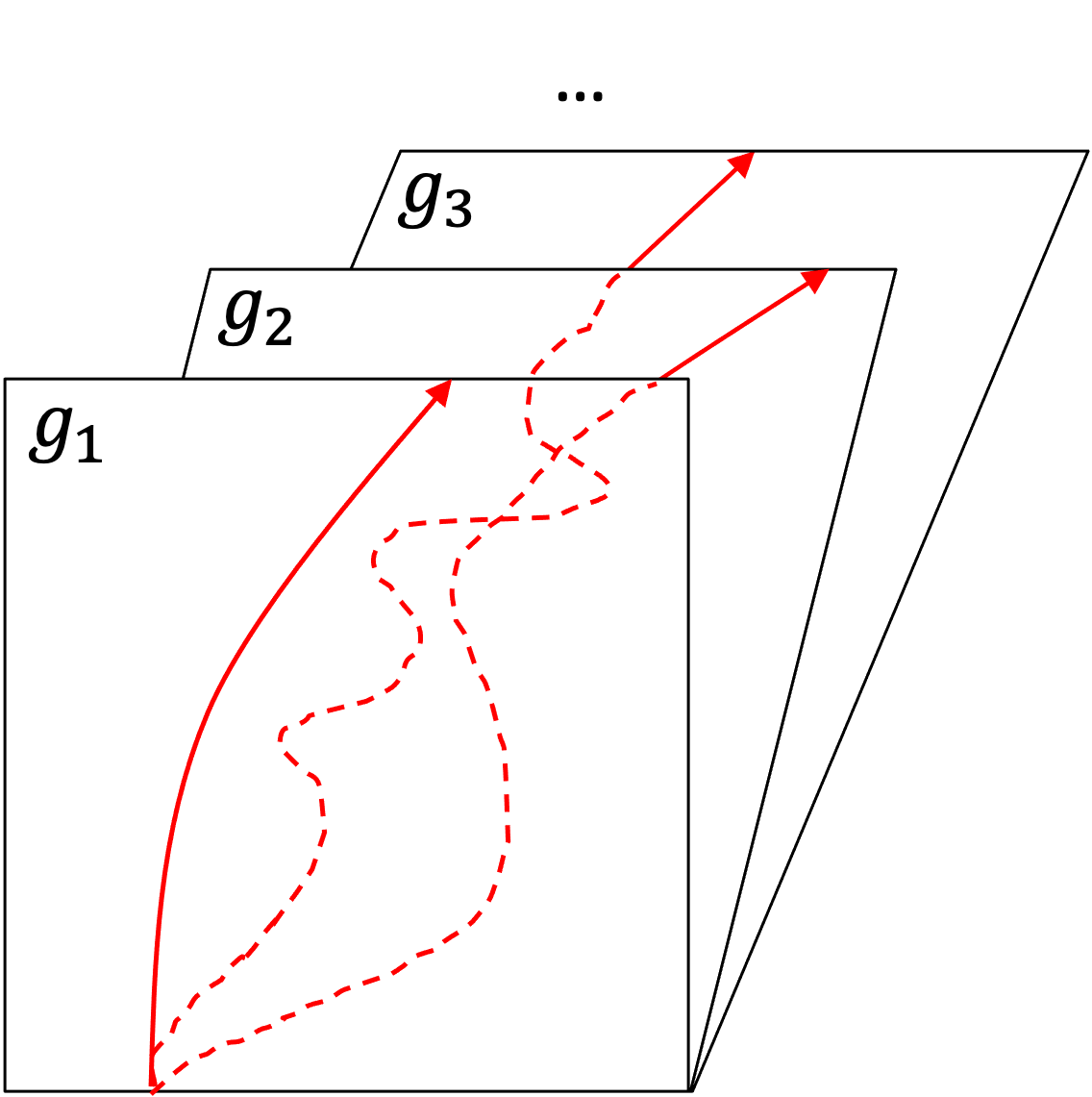}
    \caption{Light rays that enter a quantum region of spacetime at a certain location can exit at different locations in different spacetime configurations.}
    \label{fig:superpose_box}
\end{figure}

The two challenges are not disconnected, since taking the lattice refinement limit usually requires the computation of some physical quantity \cite{Ambjorn2020RenormalizationGeometryb, Steinhaus2020CoarseReviewb}. 
% (which need not always have clear observational meaning)
As pointed out in \cite{Jia2022LightGravity}, one such computable quantity is the light ray fluctuation probabilities. 
Light rays that enter a quantum region of spacetime at a certain location can exit at different locations in different spacetime configurations (\Cref{fig:superpose_box}). When there is a measurement for the exiting location, the probability distribution over the exiting locations can be computed in a Lorentzian gravitational path integral by summing over the configurations that yield the same locations. How the distribution changes as the lattice is varied offers clues to the refinement limit of the model.

In this work, we continue the study \cite{Jia2022LightGravity} of light ray fluctuation in Lorentzian simplicial quantum gravity \cite{Jia2022ComplexProspects} to consider refined lattices. While the previous study is limited to a crude symmetry-reduced box lattice with just one dynamical degree of freedom, here we refine the box lattice to incorporate as many as $2^{14}=16384$ dynamical degrees of freedom. 

A main difficulty in computing Lorentzian path integrals with many degrees of freedom in full generality is the so-called numerical sign problem. For an Euclidean path integral with positive amplitudes, the amplitudes of different configurations simply add up to positive numbers. Monte Carlo simulations are practical for such path integrals. In contrast, for a Lorentzian path integral, two complex amplitudes with large magnitudes can add up to a number with small magnitude, because the complex amplitudes can have opposite or close to opposite phase angles. Such cancellations pose higher precision requirements and render Monte Carlo simulations less practical in computing Lorentzian path integrals with many degrees of freedom.\footnote{However, see \cite{Jia2022ComplexProspects} for a Monte Carlo simulation method that overcomes the numerical sign problem for a small number of dynamical variables. See also \cite{ItoTensorCalculus} for a recent development using tensor network renormalization group methods.} We avoid this difficulty in a simplified setting, where certain lattice edge variables are non-dynamical, while the rest are dynamical but take the same value within small blocks of edges. Effectively, this amounts to composing many symmetry-reduced box models to form a larger refined box, as suggested in \cite{Jia2022LightGravity}. Since computations of the elementary boxes with few degrees of freedom can be performed efficiently through direct numerical integration, the numerical sign problem is sidestepped. As explained in \Cref{sec:rb}, this refined box model is expected to yield relevant results when the flat spacetime configuration dominates the path integral, which happens when the boundary condition is flat and when the Einstein-Hilbert term dominates the action.

Within this simplified setting, light ray fluctuation probabilities are computed numerically in 3D and 4D for a range of values of the Einstein-Hilbert term prefactor $k$. The numerical results suggest that the probabilities converge upon lattice refinement for $k$ ranging from $k=0$ to at least $k\sim 100$. Moreover, for this range of $k$ the probabilities seem to converge towards a common distribution, suggesting the presence of a universality class. A comparison across lattices indicate that while lattice refinement suppresses light ray fluctuations for some values of $k$, it enhances them for some other values of $k$ to drive both very wide and very narrow probability distributions towards a intermediate ones.

These results from the still simplified model offer encouraging hints at the existence of lattice refinement fixed points for the full theory of Lorentzian simplicial quantum gravity. As discussed in \Cref{sec:d}, with some further work the universality class may possibly be proven analytically using scaling relations. Moreover, within the simplified setting the $k=0$ case with trivial action approaches the same probability distributions as the non-trivial cases. Since the $k=0$ case does not suffer from the numerical sign problem, this suggests for future work an investigation of the $k=0$ case in a setting with fewer simplifications, given the possibility that it yields universal results relevant to other non-trivial parameters.

The paper proceeds as follows. In \Cref{sec:lsqg} we briefly summarise the relevant aspects of Lorentzian simplicial quantum gravity. In \Cref{sec:rbm} we present the refined box model, and the procedure for computing light ray fluctuation probabilities encoded in the light probability matrices. In \Cref{sec:uc} we present the numerical results, which suggest the presence of a universality class for light probability matrices. In \Cref{sec:d}, we discuss ideas towards further analytic and numerical studies.

\section{Lorentzian simplicial quantum gravity}\label{sec:lsqg}

\begin{figure}%[h]
    \centering
    \includegraphics[width=.8\textwidth]{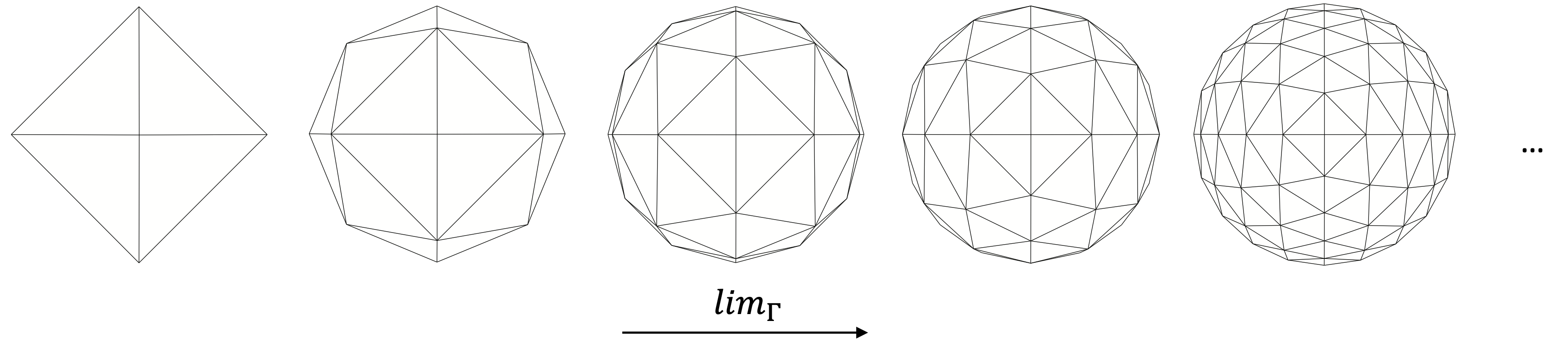}
    \caption{Lattice refinement limit.}
    \label{fig:lrl}
\end{figure}

Lorentzian simplicial quantum gravity grew out of a recent attempt to extend the much-studied approach of Euclidean simplicial quantum gravity \cite{Rocek1981QuantumCalculus, Williams1992ReggeBibliography, Loll1998DiscreteDimensions, Hamber2009QuantumApproach, Barrett2019TullioGravity} to the Lorentzian case. 
The basic idea is to give a non-perturbative definition to the formal gravitational path integral expression
\begin{align}\label{eq:qgl}
Z=&\int \mathcal{D}g ~ A[g]
\end{align}
by enumerating over piecewise flat triangulated spacetime configurations (\Cref{fig:lrl}).

While there are variations of the theory that fixes the causal signatures of the lattice edges \cite{Tate2011Fixed-topologyDomain, Tate2012Realizability1-simplex}, change to area as the basic variable \cite{Asante2021EffectiveGravity}, or enforce irregular lightcone structures \cite{Dittrich2022LorentzianSimplicial, AsanteComplexCosmology}, here we adopt the ``vanilla model'' of \cite{Jia2022ComplexProspects, Jia2022Time-spaceGravity} that takes squared length as the basic variable, and does not fix the causal signatures of the lattice edge or enforce irregular lightcone structures. In this section we very briefly summarise the relevant results, and refer the readers to \cite{Jia2022ComplexProspects, Jia2022Time-spaceGravity} for details.

The gravitational path integral is given by
\begin{align}\label{eq:pf}
Z =& \int \mathcal{D}\sigma ~ e^{E[\sigma]},
\end{align}
where the form of the path integral exponent $E[\sigma]$ is specified below. The measure $\mathcal{D}\sigma$ sums over Lorentzian piecewise flat triangulated spacetime configurations through
\begin{align}\label{eq:sqgm1}
\int \mathcal{D}\sigma =&  \lim_\Gamma\prod_{e\in\Gamma} \int_{-\infty}^\infty d\sigma_e ~ L[\sigma] C[\sigma] \mu[\sigma].
\end{align}
Here a $\Gamma$ is a finite simplicial lattice graph, i.e., the lattice graph of a finitely triangulated spacetime, and the full path integral is defined in the infinite lattice refinement limit $\lim_\Gamma$  (\Cref{fig:lrl}). The sum over Lorentzian spacetime geometries is conducted by integrating the squared lengths
% \begin{align}\label{eq:slfg}
% \sigma_e=\int_e ds^2
% \end{align}
$\sigma_e$ over all lattice edges $e$. %The sign of $\sigma_e$ indicates causal signature. 
In terms of the geodesic distance $s_e\ge 0$ along the edge, 
\begin{align}
\sigma_e=+s_e^2,0,-s_e^2
\end{align}
respectively for spacelike, lightlike, and timelike edges. Under the assumption that each simplex is a flat region of spacetime, the set $\{\sigma_e\}_{e\in\Gamma}$ of squared lengths over all edges uniquely determines the Lorentzian piecewise flat geometry.

The Lorentzian and lightcone constraints
\begin{align}\label{eq:lcd}
L[\sigma]=&
\begin{cases}
1, \quad \text{all simplices are Lorentzian},
\\
0, \quad \text{otherwise}.
\end{cases}
\\
C[\sigma]=&
\begin{cases}\label{eq:lccd}
1, \quad \text{all interior points have two lightcones attached},
\\
0, \quad \text{otherwise}.
\end{cases}
\end{align}
ensures that only Lorentzian geometries with regular lightcone structures are included in the sum. The Lorentzian constraint can be checked by computing the Cayley-Menger determinants for the (sub)simplices \cite{Jia2022ComplexProspects}. The lightcone constraint at a point can in general be checked by solving for the lightlike geodesics emanating from the point and see that they group into two cones, although in special cases the real part of the Lorentzian angles provides a shortcut to determine the number of lightcones. See \cite{Jia2022ComplexProspects, Jia2022LightGravity} (particularly Proposition 6 of \cite{Jia2022ComplexProspects}, and Sections 4.1, 5.1, 6.1 of \cite{Jia2022LightGravity}) for details. 
%In the lightcone constraint \eqref{eq:lccd}, ``interior points'' refer to all points of the simplicial complex that are not boundary points. Note that a simplex is the convex hull of its vertices, so the simplicial complex contains both points on the lattice edges, and points away from the lattice edges. Because each simplex is a flat spacetime region, each point inside the region always has two lightcones attached. Therefore the lightcone constraint needs only be checked on the boundary points of the individual simplices (which also lie in the interior of the simplicial complex). In 2D, these are points of the lattice vertices. In 3D, these are points of the lattice vertices and edges. In 4D, these are points of the lattice vertices, edges, and triangles etc \cite{Jia2022ComplexProspects, Jia2022LightGravity}. For these points, the lightcone constraint can be checked by solving for the lightlike geodesics emanating from the point and see that they group into two cones. The box model studied here (see \Cref{sec:rbm} below) is built by gluing elementary boxes studied in \Cref{Jia2022LightGravity}. The range of squared lengths meeting the lightcone constraint inside an elementary box has been determined there. Since the lightcone numbers at the boundary points of the elementary boxes agree with that of a flat box, and flat boxes glue together without affecting the lightcone constraint.

The exponent $E[\sigma]$ of \eqref{eq:pf} and measure factor $\mu[\sigma]$ of \eqref{eq:sqgm1} encode the dynamics. For reasons explained in \Cref{sec:rb}, in this work we focus on theories with just the Einstein-Hilbert term so that 
\begin{align}\label{eq:EgD}
E = & i k \sum_h \delta_h \sqrt{-\sV_h}.
\end{align}
Here $k\in\mathbb{R}$ is the coupling constant, $h$ are the hinges (codimension-2 subsimplices) of the lattice, $\delta_h$ are the deficit angle at the hinges $h$, and $\sV_h$ are the squared volume at the hinges $h$. The explicit forms of these quantities in terms of squared lengths can be found in \cite{Jia2022ComplexProspects, Jia2022Time-spaceGravity}. 
For $\mu[\sigma]$, we adopt the family of local measures\footnote{See Section 7.3 of \cite{Jia2022LightGravity} for a discussion of the physical meaning of the measure factor as accounting for the contribution from a family of configurations associated with a piecewise flat geometry.}
\begin{align}\label{eq:mf}
\mu[\sigma]=\prod_s \sV_s^{m}
\end{align}
parametrized by $m\in\mathbb{R}$ \cite{Hamber2009QuantumApproach}. Here as the product is over all simplices $s$ with squared volumes $\sV_s$. The model is therefore parameterized by the two parameters $k$ and $m$.

\section{Refined box model}\label{sec:rbm}

\subsection{Previous work}\label{sec:pw}

\begin{figure}%[h]
    \centering
    \includegraphics[width=.8\textwidth]{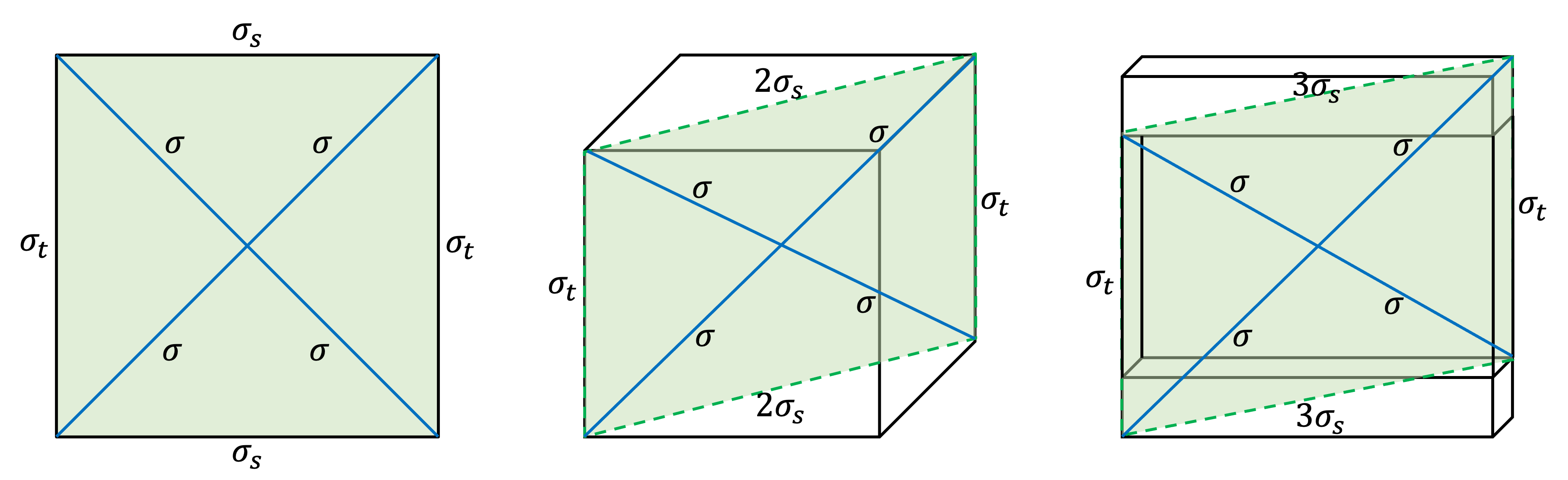}
    \caption{Symmetry-reduced box models in $D=2,3,4$ spacetime dimensions. For a generic $\sigma$, the center vertex can have non-vanishing deficit angle.}
    \label{fig:boxes}
\end{figure}

In \cite{Jia2022LightGravity} light ray fluctuation in $D=2,3,4$ spacetime dimensions is studied Lorentzian simplicial quantum gravity in symmetry-reduced box models. As illustrated in \Cref{fig:boxes}, a quantum region of spacetime is modelled as a hypercube box whose boundary edges have fixed spatial and temporal squared lengths $\sigma_s$ and $\sigma_t$. A vertex in the interior divides the box into $2D$ many hyperpyramids formed by the interior vertex as the tip and the faces of the hypercube as the bases. The model is symmetry-reduced because the interior edges are taken to share the same squared length $\sigma$. 

\begin{figure}%[h]
    \centering
    \includegraphics[width=.4\textwidth]{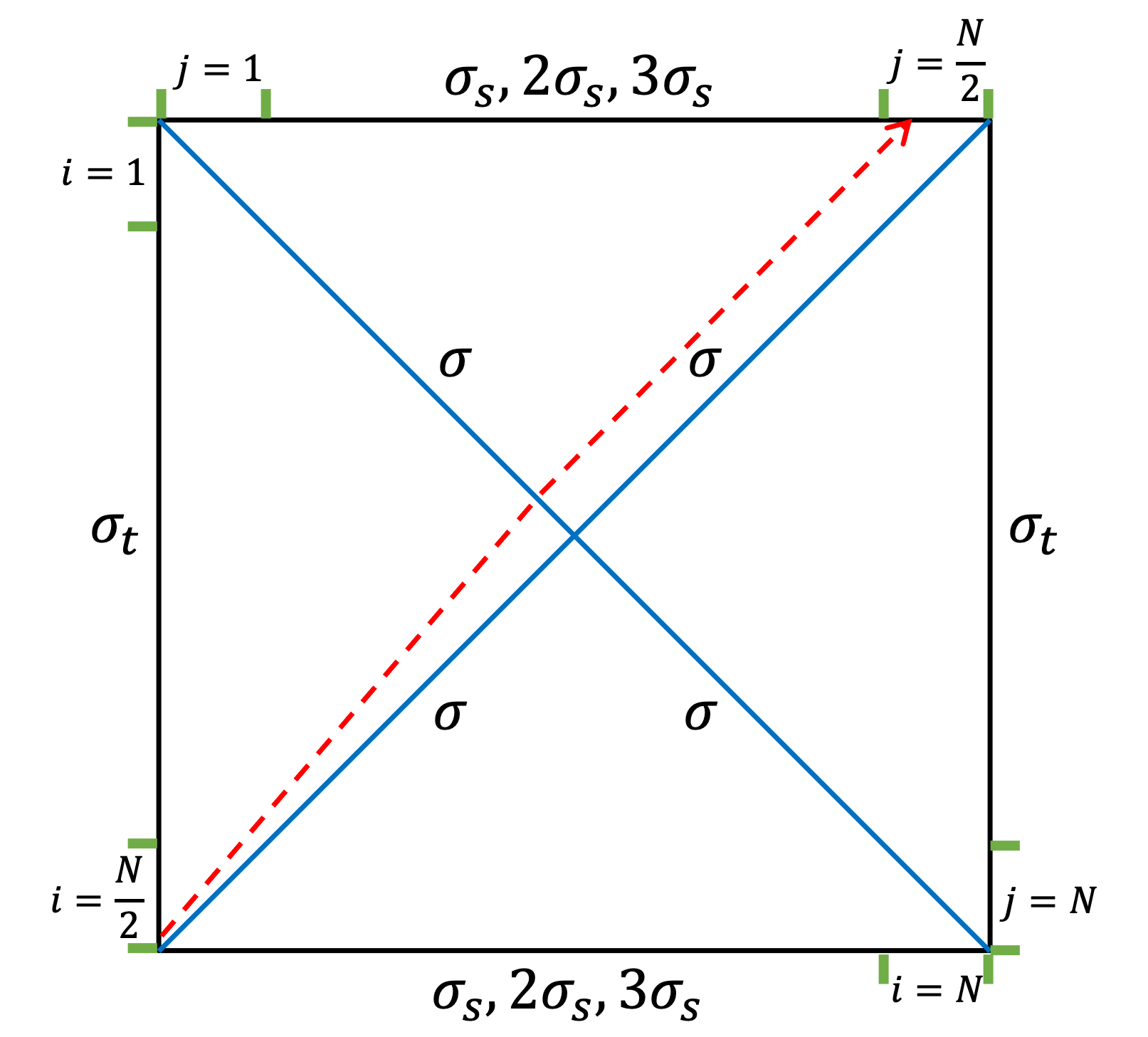}
    \caption{Light ray propagation in the diagonal plane of the boxes. The spatial boundaries in $D=2,3,4$ respectively have squared lengths $\sigma_s,2\sigma_s,3\sigma_s$.}
    \label{fig:single_box}
\end{figure}

For light rays travelling within the diagonal planes shaded in \Cref{fig:boxes}, the situation is shown in \Cref{fig:single_box}. In $D=2,3,4$, the spatial boundaries have squared lengths
\begin{align}
\sigma_s,2\sigma_s,3\sigma_s
\end{align}
since these are the diagonal lines of the hypercube faces. In the 2D diagonal plane, a test light ray either starts right-moving or left-moving. Since the whole setup is symmetric under reflection, without loss of generality we consider right-moving light rays such as the dashed one in \Cref{fig:single_box} so that they enter the box from the left or bottom side. For any given interior squared length $\sigma$ and any given in-location, the out-location can be computed by the Lorentzian trigonometry detailed in \cite{Jia2022LightGravity}, and it turns out the light rays always exit from the top or right side for the symmetry-reduced model. In the path integral $\sigma$ is integrated over, so different out-locations are set in superposition, resulting in light ray fluctuations.

To enable numerical computations for the probabilities of light ray locations, each of the four sides of the 2D plane that the light ray travels in (shaded plane of \Cref{fig:boxes}) is divided into $N/2\in \mathbb{N}$ many equally spaced intervals. The central object of interest is the \textbf{light probability matrix}
\begin{align}
p(j|i),
\end{align}
which is a $N\times N$ stochastic matrix of the conditional probabilities for light rays entering through the left-bottom in-interval $i\in \{1,2,\cdots, N\}$ to exit through the top-right out-interval $j\in \{1,2,\cdots, N\}$ (\Cref{fig:single_box}).

In \cite{Jia2022LightGravity} the light probabilities for in-location fixed at the left-bottom corner are computed in $D=2,3,4$ spacetime dimensions for a variety of model parameters by numerical integration. The results are as expected. Large light ray fluctuations in quantum regimes become suppressed as the coupling constants are tuned towards the $\hbar\rightarrow 0$ classical regime.

\subsection{Refining the boxes}\label{sec:rb}

\begin{figure}%[h]
    \centering
    \includegraphics[width=.6\textwidth]{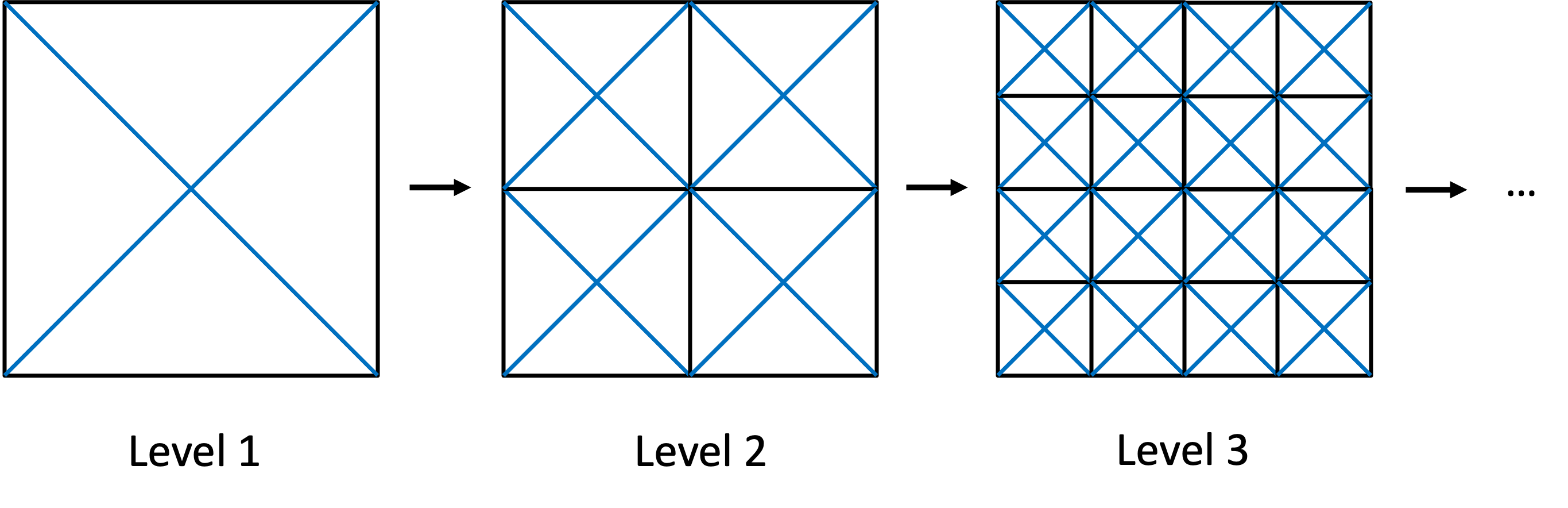}
    \caption{Refining the boxes.}
    \label{fig:refine_box}
\end{figure}

The above symmetry-reduced model is quite simplified as it incorporates only one dynamical degrees of freedom for the shared squared length at the interior edges. Although it offers reasonable qualitative results, a more detailed quantitative analysis calls for more dynamical degrees of freedom. In this work we improve the model by a systematic level-by-level refinement of the box. As illustrated in \Cref{fig:refine_box}, for each level of refinement we replace an old box by composing four elementary boxes half the length dimension of the old box. Level 1 refinement is the original model, level 2 refinement includes $2^2=4$ boxes, level 3 refinement includes $2^4=16$ boxes etc. In general, level $L$ refinement includes $2^{2L-2}$ boxes.

Ideally, we would like to treat all the interior horizontal, vertical, and diagonal edges as dynamical, and take the infinite lattice refinement limit. Practically, this is unfeasible since a numerical computation is always performed on a finite lattice, and treating all (as many as $2^{14}=16384$ in the model studied below) interior edges as dynamical is beyond our current technical capability due to the numerical sign problem for Lorentzian path integrals. Therefore, we make two simplifications:
\begin{itemize}
    \item Frozen frame. We only treat the diagonal interior edges as dynamical, but keep fixed all the interior horizontal and vertical edges. This sets up a rigid frame for the elementary box boundaries, but allows their interior geometries to fluctuate.
    \item Partial symmetry. We take the diagonal interior edges within the same elementary box to share the same squared length, while allowing differences across different boxes. This imposes a partial symmetry-reduction limited within the elementary boxes.
\end{itemize}
In effect, we model a big box as composed of many identical elementary symmetry-reduced boxes defined in \Cref{sec:pw} to allow many dynamical degrees of freedom in the interior of the big box. Computations for the big box are enabled by composing results from the elementary boxes.

% Another way to understand the simplification is to start with a flat spacetime configuration on a refined lattice in \Cref{fig:refine_box}. The ``frozen frame'' assumption amounts to keeping the squared lengths at the horizontal and vertical edges fixed at the flat configuration values, but allowing the squared lengths at the diagonal edges to vary. The ``frozen frame'' assumption amounts to setting the diagonal edges in the same elementary box to share the same squared length. 

These simplifications are suitable when the flat spacetime configuration dominates the path integral. This happens when the action for the path integral contains just the Einstein-Hilbert term, and the boundary condition is that of flat spacetime (however, see \Cref{sec:liff} for a discussion on a limitation of the rigid frame assumption). Under this condition, the flat spacetime configuration constitutes the saddle point of the action, so the phase of the amplitude varies slowly around the flat configuration, which allows it to dominate the path integral. In this case, the lengths of the rigid frame edges are set to agree with those of the flat configuration, and the dynamical edges within the elementary boxes incorporate quantum fluctuations around the flat configuration. 

Therefore, in the following we focus on models defined in \Cref{sec:lsqg} with just the Einstein-Hilbert term, and pick flat boundary conditions.

\subsection{Procedure for computing the light probability matrices}\label{sec:pclpm}

To obtain the refinement level $L$ light probability matrix, we first compute the light amplitude matrix
\begin{align}\label{eq:lam}
A_{i,j}:=\int_{D[i,j]} \mathcal{D}\sigma ~ e^{E[\sigma]}
\end{align}
for its elementary box without refinement. Here $D[i,j]$ for $i,j\in \{1,2,\cdots, N\}$ is the domain where light rays entering through the in-interval $i$ exit through the out-interval $j$ (\Cref{fig:single_box}). In a concrete computation we set the light ray to always come in through the middle of the in-intervals. This \textit{ad hoc} choice does not matter much in the refinement limit, where the interval becomes small. By the Lorentzian trigonometry method of Section 3 in \cite{Jia2022LightGravity}, the out-location as a function of the interior squared length (recall that the elementary box has all its interior squared lengths identified) can be solved, and is seen to be monotonic. Therefore, the domains $D[i,j]$ for different $j$ can be determined numerically by inverting this function.

Given the domains $D[i,j]$, the one-dimensional integrals \eqref{eq:lam} could be computed by direct numerical integration, and we call this the ``interval integration'' method. In advanced numerical integration algorithms that check against error bounds, this method converges within a reasonable amount of time only when the phase of the integrand does not vary too fast, which is not always the case. A complementary alternative is the ``midpoint approximation'' method, where we pick the midpoint $z=(x+y)/2$ of the integration domain $(x,y)$ and approximate the integral $\int_x^y f(\sigma)d\sigma$ by $(y-x)f(z)$. This method always yields a result within a reasonable amount of time, even when the integrand varies fast, but the result is not guaranteed to be accurate. %In particular, in evaluating $\sum_{n=1}^N A_{p,q}$ that show up in expressions such as \eqref{eq:aup} and \eqref{eq:alw}, while the exact result is independent of the choice of $m$, that for the midpoint approximation method depends on it due to the inaccuracy. To fix the convention we pick $m=N/2$.
In the computations presented in \Cref{sec:uc} we will apply both methods and compare the results when possible. We will see that when both methods are applicable, they yield close results at high refinement levels. %This suggests that although midpoint approximation is not the most accurate method, it is of use for parameter ranges where interval integration is inapplicable.

\begin{figure}%[h]
    \centering
    \includegraphics[width=.4\textwidth]{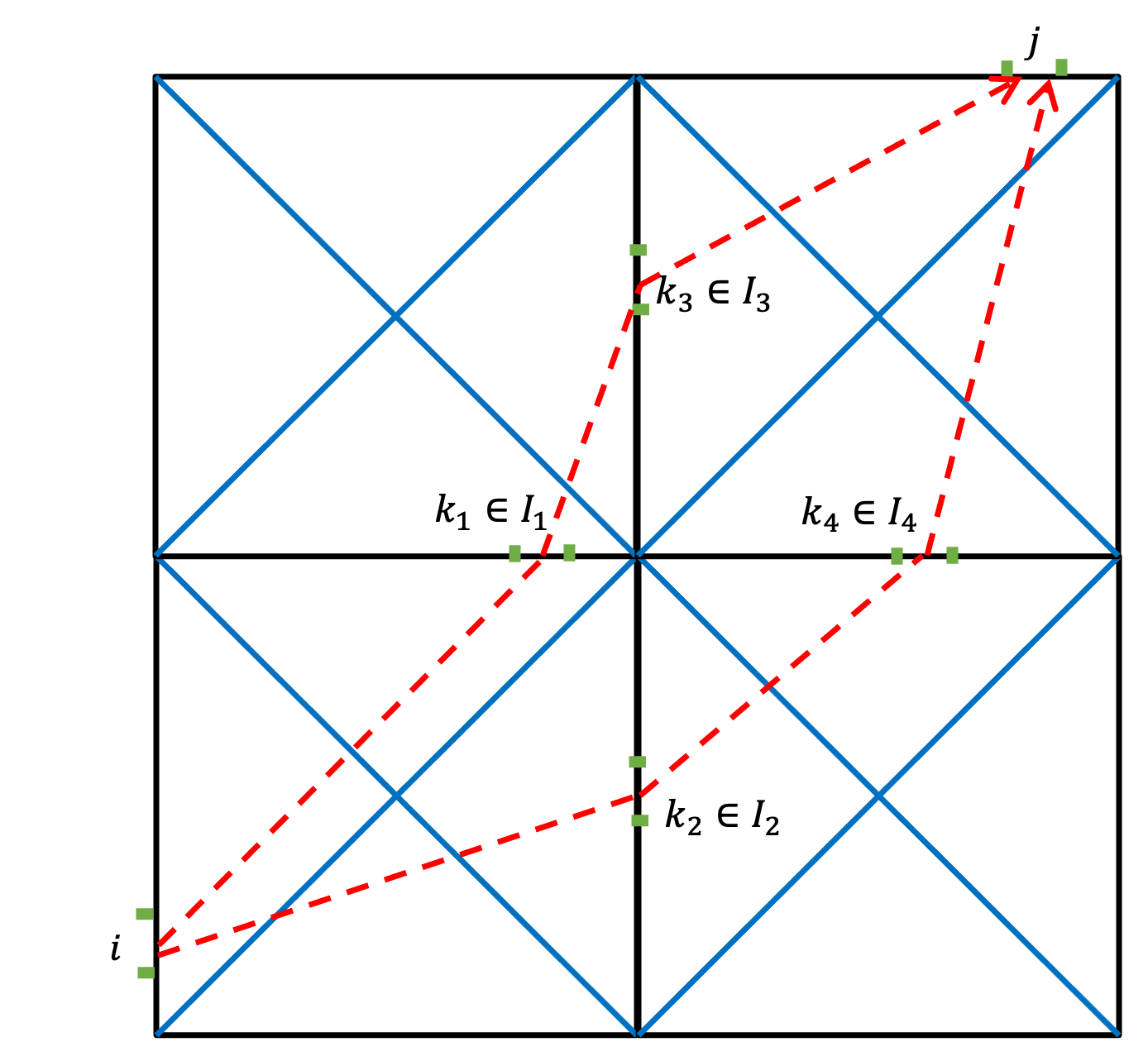}
    \caption{Alternative light paths leading to the same pair of in- and out-intervals.}
    \label{fig:compose_box}
\end{figure}

The next step is to compose the light amplitude matrix \eqref{eq:lam} of the $2^{2L-2}$ elementary boxes to obtain the light amplitude matrix for the largest box. Since the right-moving light rays we consider always move from left/bottom to top/right across an elementary box, they do the same across any larger box. According to quantum theory, for each fixed in-out interval pair $i,j\in \{1,2,\cdots, N 2^{L-1}\}$ for the largest box, we need to sum over all interior configurations that allow light rays entering at $i$ to exit at $j$. In practise, computational complexity can be significantly reduced by a Kadanoff-type blocking procedure. At each step, we block four old small boxes into a new box, whose light amplitude matrix is obtained by summing over the products of all the relevant light amplitudes for the old boxes. 

For instance, $i$ and $j$ in \Cref{fig:compose_box} are connected by light rays through either the ``upper'' or the ``lower'' route shown in the figure. For $A_{p,q}$ as the amplitudes of the old boxes, these routes have amplitudes 
\begin{align}\label{eq:aup}
A'_{upper}=&\sum_{k_1\in I_1, k_3\in I_3}A_{i,k_1}A_{k_1,k_3}A_{k_3,j}\sum_q A_{p,q},
\\
A'_{lower}=&\sum_{k_2\in I_2, k_4\in I_4}A_{i,k_2}A_{k_2,k_4}A_{k_4,j}\sum_q A_{p,q}.
\end{align}\label{eq:alw}
Here the $\sum_q A_{p,q}$ term refers to the box that the light ray does not go through. We sum over spacetime configurations compatible with all would be out-intervals $q$ in this box, because as the light ray does not actually go through it, all of the configurations are compatible with the actual in-interval $i$ and out-interval $j$ for the larger box. The in-interval $p$ is arbitrary since they all lead to the same result, which is \eqref{eq:lam} integrated over all $\sigma$. Although the results from the interval integration method implement this well, the results from the midpoint approximation method do exhibit differences for different $p$ on coarse partitions with small $N$. The difference becomes small as the refinement level goes up, but to reduce the ambiguity we pick $p=N/2$ in \eqref{eq:aup} and \eqref{eq:alw}. Altogether, the $i,j$ element for the light amplitude matrix of the larger box is 
\begin{align}
A'_{i,j}=A'_{upper}+A'_{lower}.
\end{align}
All other elements can be obtained in a similar way. 

Schematically, each step of blocking reverses the arrow of \Cref{fig:refine_box} to reduce the total number of boxes by four. Iterating this will eventually yield the light amplitude matrix $A_{i,j}$ for the largest box with $N 2^{L-1}\times N 2^{L-1}$ entries. Then
\begin{align}
p(j|i)=\abs{A_{i,j}}^2/\sum_j \abs{A_{i,j}}^2
\end{align}
yields the normalised probabilities for light ray propagation across the total region. 

To compare $p(j|i)$ at different refinement levels, we can coarsen $p(j|i)$ by partitioning the in and out intervals $\{1,2,\cdots, N 2^{L-1}\}$ and summing over $p(j|i)$ within the intervals. For instance, to compare with $N\times N$ light probability matrices at level $L=1$, we can partition the in- and out-intervals into $N$ subsets $S_1=\{1,2\cdots,2^{L-1}\}$, $S_2=\{2^{L-1}+1,2^{L-1}+2\cdots,2^{L-1}+2^{L-1}\}$ etc., and take sums to obtain the coarse $N\times N$ matrix
\begin{align}
p(J|I)=&\sum_{i\in S_I,j\in S_J}p(j|i)/\sum_{i\in S_I,j}p(j|i)
\\
=&\sum_{i\in S_I,j\in S_J}p(j|i)/\abs{S_I},
\end{align}
where $\abs{S_I}$ is the cardinality of $S_I$. If as the refinement level goes up the coarse $N\times N$ matrices $p(J|I)$ converge towards a fixed point, the result can be taken to hold on an infinitely refined lattice. The numerical results on finite lattices presented next indeed hint at such convergence.

\section{Universality class?}\label{sec:uc}

\begin{figure}%[h]
    \centering
    \includegraphics[width=.7\textwidth]{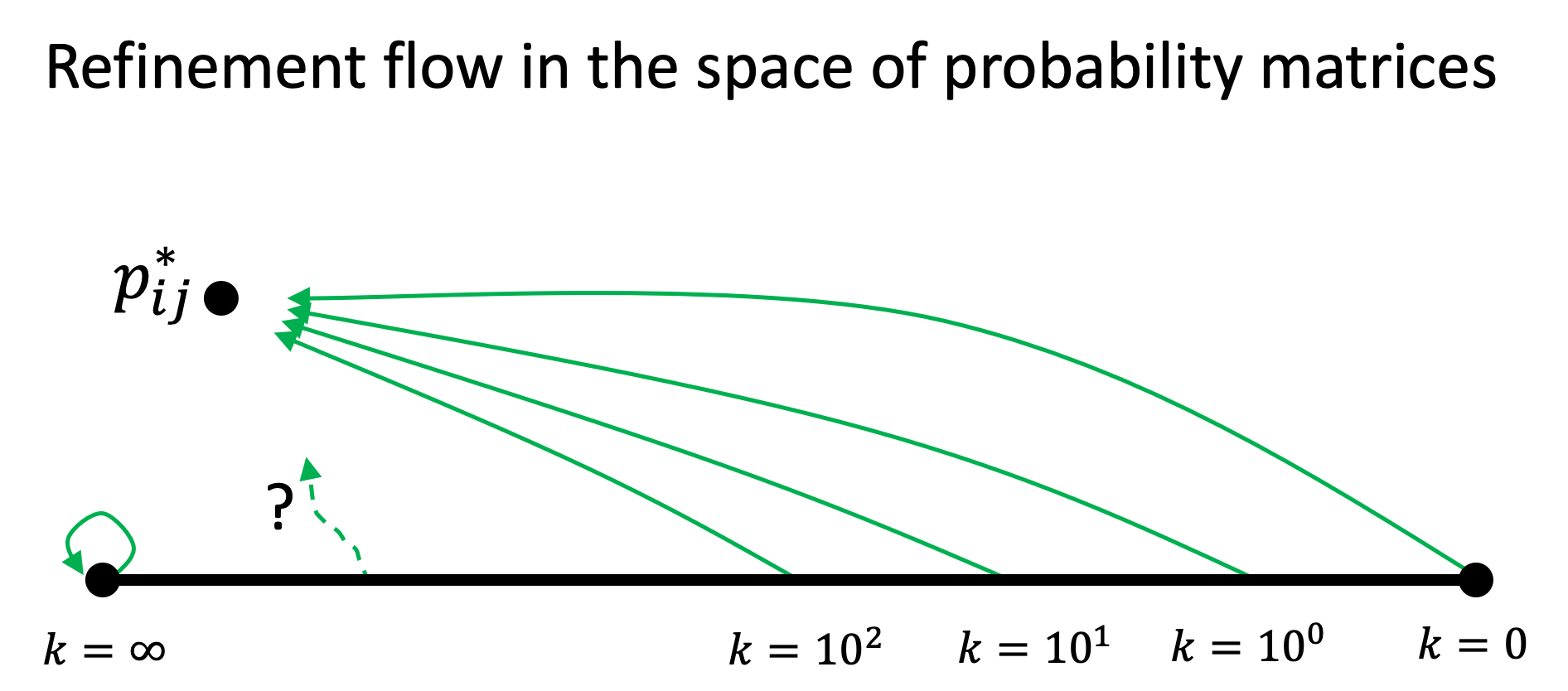}
    \caption{As the lattice is refined, the light probability matrices for a range of finite $k$ flow towards a common value. The strict classical limit $k=\infty$ is a fixed point since there is no light ray fluctuation. The intermediate values between $k\sim 100$ and $k=\infty$ is not probed due to numerical limitations. See \Cref{sec:oub} for an analytic discussion.}
    \label{fig:pmflow}
\end{figure}

Following the above procedure, we numerically compute the light probability matrices for ever higher levels of refinement and check if the results converge. The main results are summarized qualitatively in \Cref{fig:pmflow}. Very interestingly, upon refinement the light probability matrices not only show signs of convergence, but also show signs of convergence toward a common value for a range of different $k$ values. This suggests universal behaviour, indicating that in the lattice refinement limit the light probability matrices are not very sensitive to the value of the coupling constant $k$ in a large interval. 

In this section we will explain how the results are obtained at the quantitative level. In  \Cref{sec:d} we will comment further on the universal behavior.

\subsection{Numerical setup}

For the parameters of the numerical computations, we adopted spacetime dimensions $D=3,4$, refinement levels $L=1,\cdots, 8$, the number of light ray incoming/outgoing intervals $N=16,32$, and the measure factor exponent $m=0,-1/12$. Since we are interested in how the light probability matrices depend on the gravitational coupling constant $k$, we will explore various $k$ values in the computations below.

We compute up to $4$ spacetime dimensions because this is the value for our universe, and did not consider the case $D=2$ as in \cite{Jia2022LightGravity} because the Einstein-Hilbert term is a trivial topological invariant in 2D by a Lorentzian version of the Gauss-Bonnet theorem \cite{SorkinLorentzianVectors}. The refinement level is taken up to $L=8$, and the number of light ray intervals is taken up to $N=32$, both to keep the computation efficient on a personal desktop. Certainly, these parameters can be further increased with better devices, but the available results already provide strong clues for convergence and universal behavior so we stop here and suggest an analytical investigation for further studies in \Cref{sec:oub}. 
% At $L=8$ and $N=32$, we need to compose $2^{L-1}\times 2^{L-1}=2^{14}$ many elementary boxes represented by $N/2\times N/2=16\times 16$ light amplitude matrices into a giant $2^{L-1}N/2\times 2^{L-1}N/2=2^{11}\times 2^{11}$ matrix.
The values $m=0,-1/12$ for the measure factor exponent are suggested by previous studies. The DeWitt measure is for $m=-1/12$ in 3D, and $m=0$ in 4D \cite{Hamber2009QuantumApproach}. As shown in Figure 25 and Figure 37 of \cite{Jia2022LightGravity}, the only range of feasible $m$ values are closely around the above ones. Outside this range the measure term unphysically dominates over the action to enhance configurations just based on spacetime volumes and disregards curvatures. 

\begin{figure}%[h]
    \centering
    \includegraphics[width=.4\textwidth]{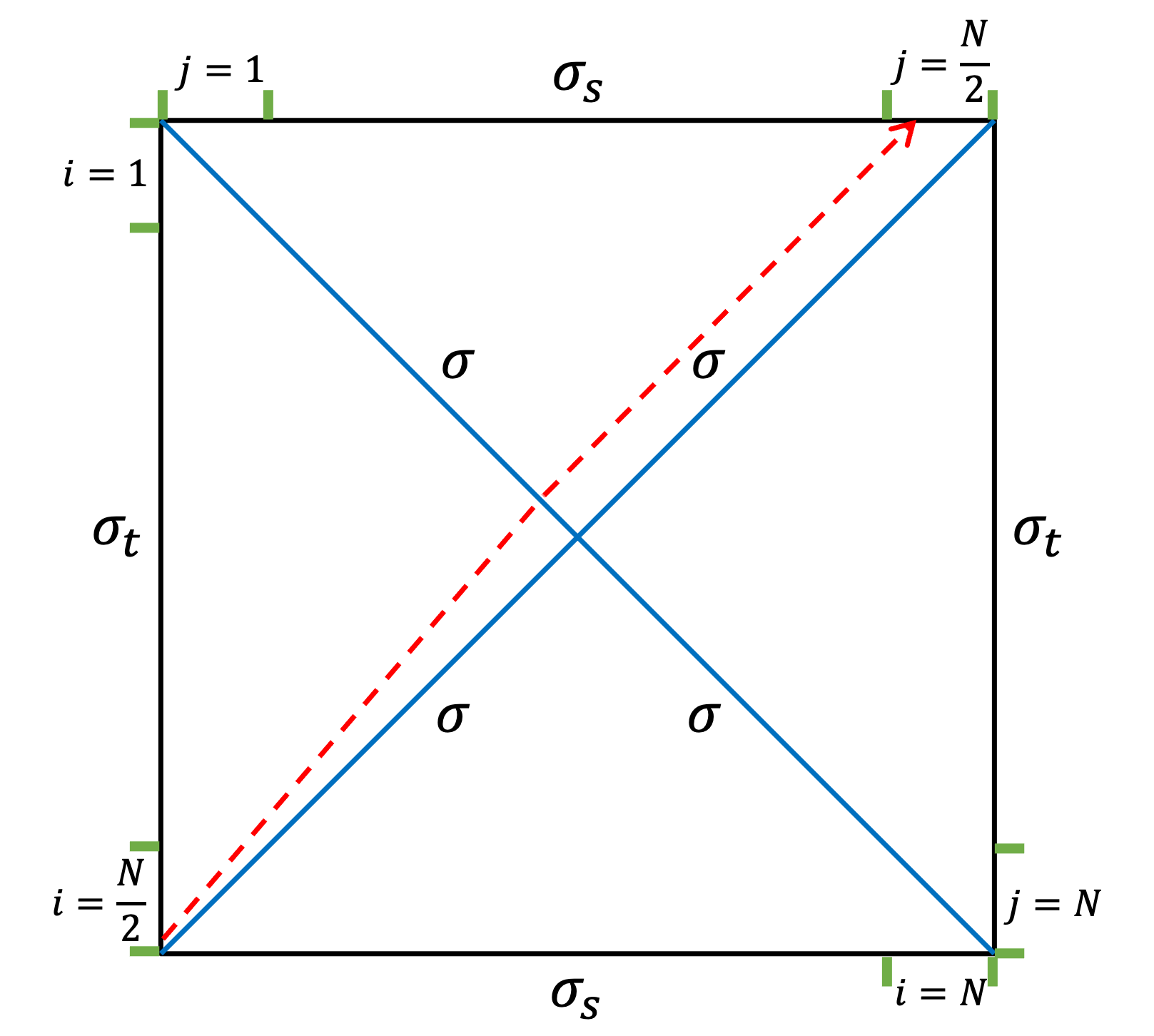}
    \caption{Light ray propagation in the diagonal plane of the boxes. The spatial boundaries have squared length $\sigma_s$ in both $D=3$ and $D=4$.}
    \label{fig:single-box2}
\end{figure}

We focus on light rays that travel within the box diagonal plane (\Cref{fig:single-box2}). For concreteness we set the diagonal plane to be ``square shaped'', i.e., to obey $\abs{\sigma_s}=\abs{\sigma_t}$ for its spatial boundary squared length $\sigma_s$ and temporal boundary squared length $\sigma_t$. To make meaningful comparison across different refinement levels we fix the value
\begin{align}
\abs{\sigma_s}=\abs{\sigma_t}=1
\end{align}
for the largest box built out of composition. Note that this setting is different from the previous work \cite{Jia2022LightGravity} where the spatial boundaries of the diagonal plane assume different values in different dimensions (\Cref{fig:single_box}).

In quantifying the difference between two light probability matrices $p,p'$, we use the measure
\begin{align}\label{eq:pdiff}
d(p,p'):=\max_{i,j}\abs{p(j|i)-p'(j|i)}.
\end{align}

\subsection{Results in 3D}

\subsubsection{N=16}

Let us first consider $N=16$, and compare the interval integration and midpoint approximation methods mentioned in \Cref{sec:rbm}. \Cref{tab:table-0} shows that when both methods are efficiently applicable, they yield close results at the highest refinement level. However, as $k$ gets large, the integrand phase oscillates fast to the extent that the interval integration method  fails to converge in a reasonable amount of time. Therefore in the following data tables we show results from the midpoint approximation method because it yields results for a wider range of parameters.

\begin{table}[H]
\caption{\label{tab:table-0}Interval integration vs. midpoint approximation methods for $D=3, N=16, L=8$. The table shows $d(p,p')$ for light probability matrices $p$ and $p'$ evaluated using the two methods. At $k=1000$ the interval integration method becomes inefficient so the comparison is not made.} 
\centering
\begin{tabular}{llllll}
    \hline\hline
     & \textbf{k=0.0} & \textbf{k=1.0} & \textbf{k=10.0} & \textbf{k=100.0} & \textbf{k=1000.0} \\\hline
    m=0.0 & 0.0 & 5.3e-7 & 0.001 & 0.0053 & N/A \\
    m=-0.083 & 0.073 & 0.073 & 0.073 & 0.074 & N/A \\\hline\hline
  \end{tabular}
\end{table}

\begin{table}[H]
\caption{\label{tab:table-1}Convergence measured by $d(p_L,p_{L-1})$   as the refinement level $L$ increases for $D=3, N=16$.}
\centering
\begin{tabular}{llllllll}
    \hline\hline
    \textbf{Level L} & \textbf{2} & \textbf{3} & \textbf{4} & \textbf{5} & \textbf{6} & \textbf{7} & \textbf{8} \\\hline
    m=0.0, k=0.0 & 0.28 & 0.36 & 0.35 & 0.11 & 0.065 & 0.04 & 0.023 \\
    m=0.0, k=1.0 & 0.28 & 0.36 & 0.35 & 0.11 & 0.065 & 0.04 & 0.023 \\
    m=0.0, k=10.0 & 0.36 & 0.47 & 0.38 & 0.11 & 0.065 & 0.04 & 0.023 \\
    m=0.0, k=100.0 & 0.33 & 0.41 & 0.43 & 0.4 & 0.35 & 0.13 & 0.035 \\
    m=0.0, k=1000.0 & 0.41 & 0.45 & 0.48 & 0.23 & 0.41 & 0.42 & 0.44 \\
    m=-0.083, k=0.0 & 0.68 & 0.57 & 0.59 & 0.14 & 0.067 & 0.033 & 0.015 \\
    m=-0.083, k=1.0 & 0.68 & 0.57 & 0.59 & 0.14 & 0.067 & 0.033 & 0.015 \\
    m=-0.083, k=10.0 & 0.67 & 0.55 & 0.59 & 0.14 & 0.067 & 0.033 & 0.015 \\
    m=-0.083, k=100.0 & 0.68 & 0.64 & 0.61 & 0.5 & 0.6 & 0.05 & 0.019 \\
    m=-0.083, k=1000.0 & 0.67 & 0.74 & 0.54 & 0.41 & 0.32 & 0.32 & 0.57 \\\hline\hline
  \end{tabular}
\end{table}

\begin{table}[H]
\caption{\label{tab:table-2}For $D=3, N=16, L=8$ and $k$ values $(k_1,k_2,k_3,k_4,k_5):=(0,1,10,100,1000)$, showing $d[i]:=d(p_{k_i},p_{k_{i+1}})$. The light probability matrices for $k\in\{k_1,k_2,k_3,k_4\}$ are very close at this level.}
\centering
  \begin{tabular}{lllll}
    \hline\hline
     & \textbf{d[1]} & \textbf{d[2]} & \textbf{d[3]} & \textbf{d[4]} \\\hline
    m=0.0 & 9.4e-7 & 0.00011 & 0.01 & 0.47 \\
    m=-0.083 & 3.5e-7 & 3.5e-5 & 0.0036 & 0.25 \\\hline\hline
  \end{tabular}
\end{table}

\begin{figure}[H]
    \centering
    \includegraphics[width=1.\textwidth]{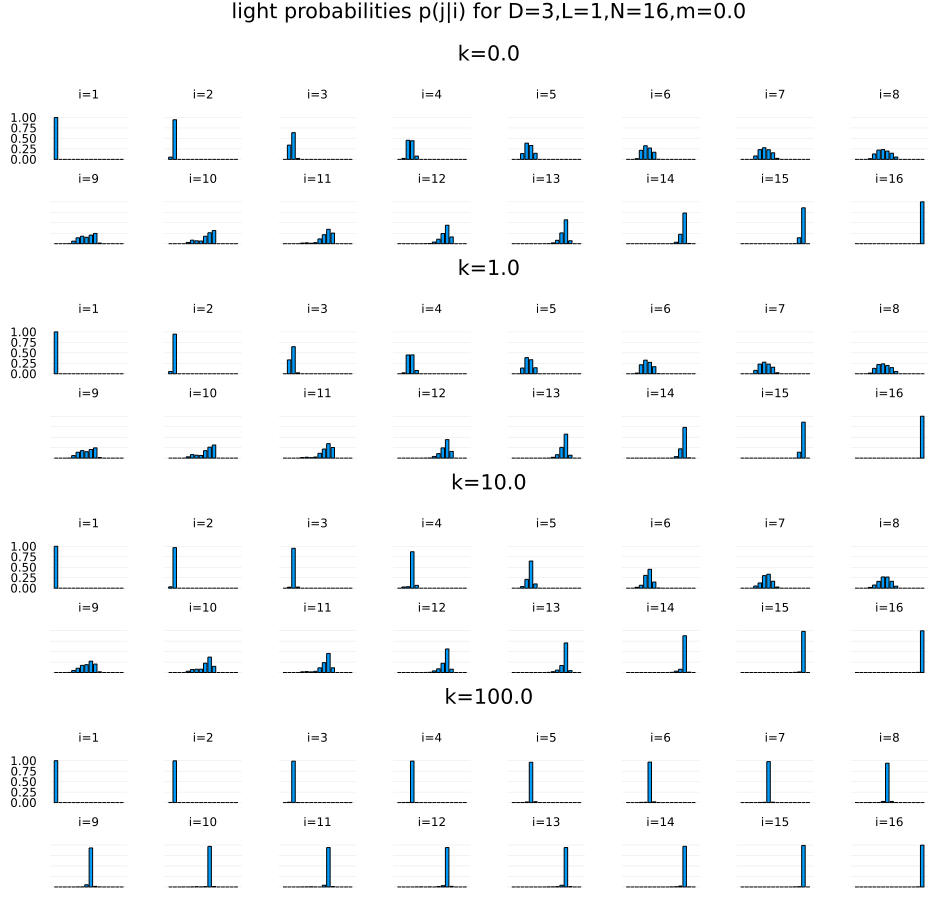}
    \caption{Initial light probability matrices without refinement as evaluated in the interval integration method. The horizontal axis enumerates different $j$, the outgoing intervals (the same applies to the figures of $p(i|j)$ below). As $k$ is increased the probabilities are more sharply peaked.}
    \label{fig:probMat_D3-L1-N16-m0}
\end{figure}

\begin{figure}[H]
    \centering
    \includegraphics[width=1.\textwidth]{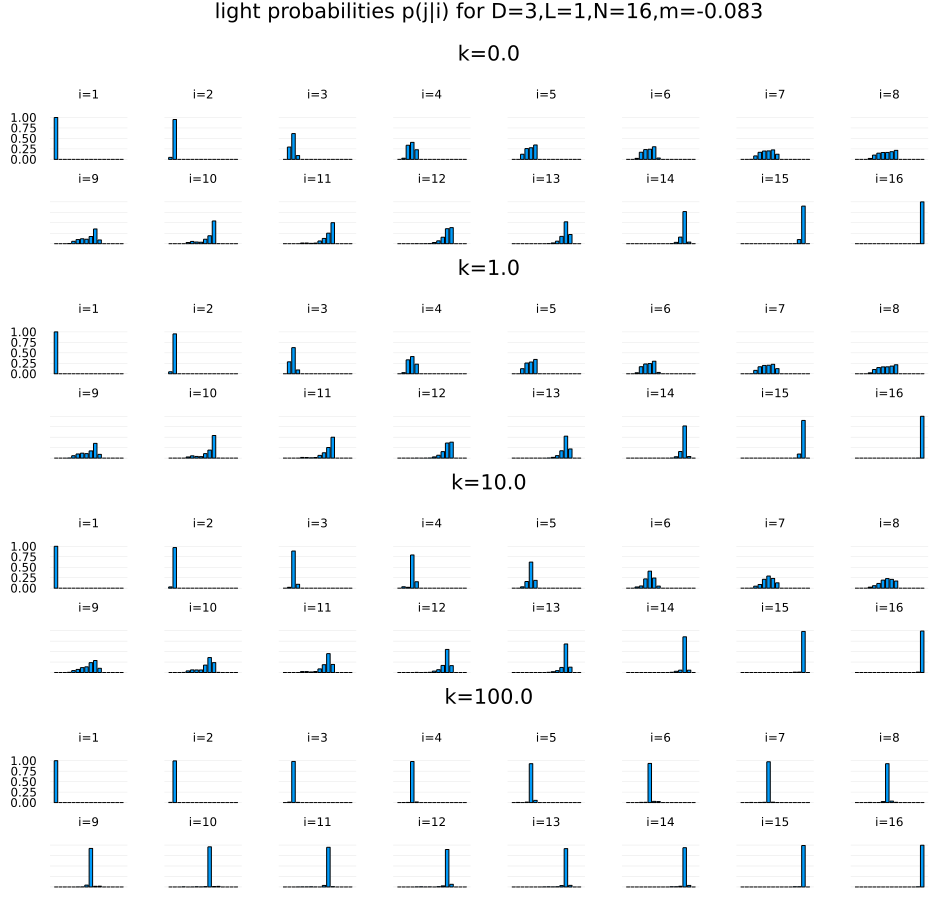}
    \caption{Initial light probability matrices without refinement as evaluated in the interval integration method. As $k$ is increased the probabilities are more sharply peaked.}
    \label{fig:probMat_D3-L1-N16-m-}
\end{figure}

\begin{figure}[H]
    \centering
    \includegraphics[width=1.\textwidth]{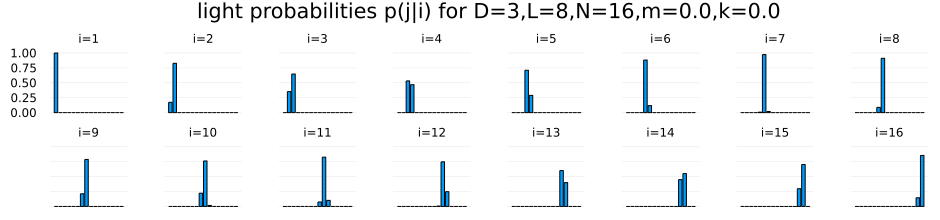}
    \caption{The commonly approached light probability matrix for $N=16,m=0$.}
    \label{fig:probMat_D3-L8-N16-m0}
\end{figure}

\begin{figure}[H]
    \centering
    \includegraphics[width=1.\textwidth]{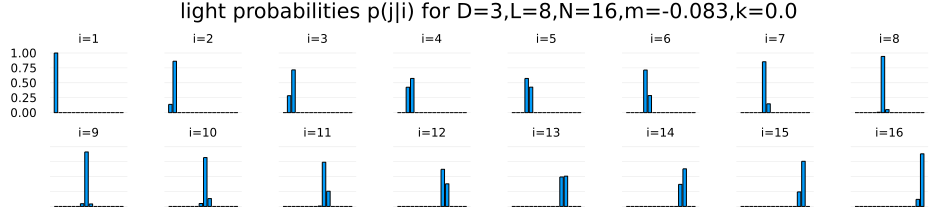}
    \caption{The commonly approached light probability matrix for $N=16,m=-1/12$.}
    \label{fig:probMat_D3-L8-N16-m-}
\end{figure}

For $k$ values $(k_1,k_2,k_3,k_4)=(0,1,10,100)$, the light probability matrices without refinement exhibit different shapes (\Cref{fig:probMat_D3-L1-N16-m0} for $m=0$; \Cref{fig:probMat_D3-L1-N16-m-} for $m=-1/12$). For each individual $k$ value,  as the lattice is refined the light probability matrices show signs of convergence as measured by $d(p_L,p_{L-1})$    for adjacent refinement levels $L$ and $L-1$ (\Cref{tab:table-1}). Moreover, even for different $k$ values, the light probability matrices seem to converge to a common one (\Cref{fig:probMat_D3-L8-N16-m0} for $m=0$; \Cref{fig:probMat_D3-L8-N16-m-} for $m=-1/12$), as indicated by the small values of $d[i]:=d(p_{k_i},p_{k_{i+1}})$ at the highest refinement level $L=8$ for $i=1,2,3$ (\Cref{tab:table-2}). 

Comparing light probabilities at the no refinement level (\Cref{fig:probMat_D3-L1-N16-m0}, \Cref{fig:probMat_D3-L1-N16-m-}) and those at the highest refinement level (\Cref{fig:probMat_D3-L8-N16-m0}, \Cref{fig:probMat_D3-L8-N16-m-}), we see that lattice refinement cannot be said to just enhance or suppress light ray fluctuations. While for parameters such as $k=0$ with wide probability distributions, refinement tends to make them narrower, for parameters such as $k=100$ with narrow probability distributions, refinement tend to make them wider. The correct understanding seems to be that refinement drives both very wide and very narrow distributions to an intermediate one.

For $k_5=1000$, as seen from \Cref{tab:table-2} at the refinement level $L=8$ the light probability matrices have not converged. This leads to the question whether at higher refinement levels the results will converge, and whether the resulting light probability matrix is the same as before. More generally, one can ask exactly which class of $k$ values belongs to the above universality class. These questions are discussed further in \Cref{sec:oub}.

\subsubsection{N=32}

\begin{figure}[H]
    \centering
    \includegraphics[width=1.\textwidth]{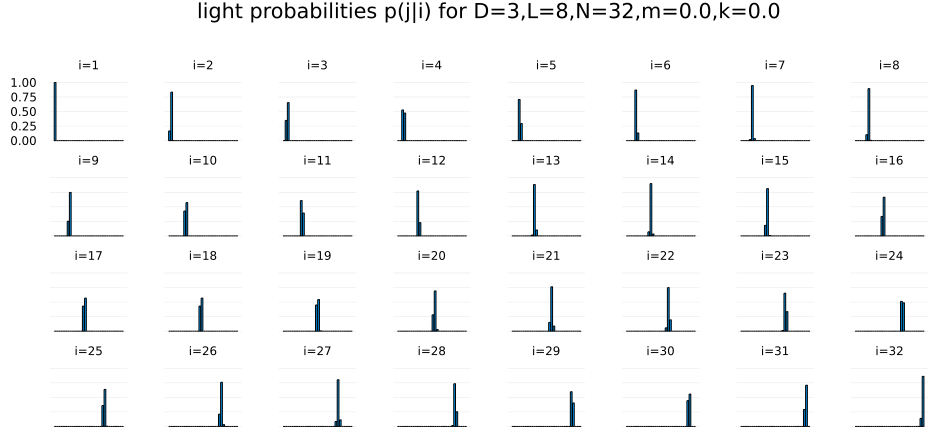}
    \caption{The commonly approached light probability matrix for $N=32,m=0$.}
    \label{fig:probMat_D3-L8-N32-m0}
\end{figure}

\begin{figure}[H]
    \centering
    \includegraphics[width=1.\textwidth]{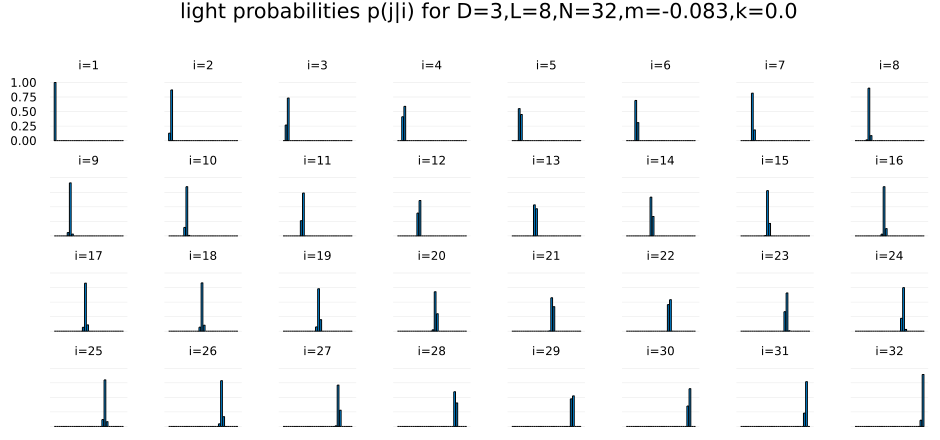}
    \caption{The commonly approached light probability matrix for $N=32,m=-1/12$.}
    \label{fig:probMat_D3-L8-N32-m-}
\end{figure}

For $N=32$ the numerical results are shown in \Cref{tab:table-0D3N32}, \Cref{tab:table-1D3N32}, \Cref{tab:table-2D3N32}, \Cref{fig:probMat_D3-L8-N32-m0}, and \Cref{fig:probMat_D3-L8-N32-m-}. The results are qualitatively similar to the $N=16$ case, so we will not comment further on them.

\begin{table}[H]
\caption{\label{tab:table-0D3N32}Interval integration vs. midpoint approximation methods for $D=3, N=32, L=8$. The table shows $d(p,p')$ for light probability matrices $p$ and $p'$ evaluated using the two methods. At $k=1000$ the interval integration method becomes inefficient so the comparison is not made.} 
\centering
  \begin{tabular}{llllll}
    \hline\hline
     & \textbf{k=0.0} & \textbf{k=1.0} & \textbf{k=10.0} & \textbf{k=100.0} & \textbf{k=1000.0} \\\hline
    m=0.0 & 0.0 & 5.3e-7 & 0.001 & 0.0053 & N/A \\
    m=-0.083 & 0.039 & 0.039 & 0.039 & 0.038 & N/A \\\hline\hline
  \end{tabular}
\end{table}

\begin{table}[H]
\caption{\label{tab:table-1D3N32}Convergence measured by $d(p_L,p_{L-1})$ as the refinement level $L$ increases for $D=3, N=32$.} 
\centering
  \begin{tabular}{llllllll}
    \hline\hline
    \textbf{Level L} & \textbf{2} & \textbf{3} & \textbf{4} & \textbf{5} & \textbf{6} & \textbf{7} & \textbf{8} \\\hline
    m=0.0, k=0.0 & 0.21 & 0.3 & 0.23 & 0.23 & 0.13 & 0.1 & 0.067 \\
    m=0.0, k=1.0 & 0.21 & 0.3 & 0.23 & 0.23 & 0.13 & 0.1 & 0.067 \\
    m=0.0, k=10.0 & 0.18 & 0.3 & 0.26 & 0.23 & 0.13 & 0.1 & 0.067 \\
    m=0.0, k=100.0 & 0.46 & 0.56 & 0.71 & 0.85 & 0.66 & 0.37 & 0.071 \\
    m=0.0, k=1000.0 & 0.61 & 0.64 & 0.8 & 0.93 & 0.72 & 0.75 & 1.0 \\
    m=-0.083, k=0.0 & 0.34 & 0.39 & 0.28 & 0.23 & 0.16 & 0.11 & 0.081 \\
    m=-0.083, k=1.0 & 0.34 & 0.39 & 0.28 & 0.23 & 0.16 & 0.11 & 0.081 \\
    m=-0.083, k=10.0 & 0.25 & 0.36 & 0.26 & 0.24 & 0.16 & 0.11 & 0.081 \\
    m=-0.083, k=100.0 & 0.45 & 0.68 & 0.68 & 0.77 & 0.77 & 0.34 & 0.087 \\
    m=-0.083, k=1000.0 & 0.55 & 0.61 & 0.83 & 0.93 & 0.81 & 0.43 & 0.97 \\\hline\hline
  \end{tabular}
\end{table}

\begin{table}[H]
\caption{\label{tab:table-2D3N32}For $D=3, N=32, L=8$ and $k$ values $(k_1,k_2,k_3,k_4,k_5):=(0,1,10,100,1000)$, showing $d[i]:=d(p_{k_i},p_{k_{i+1}})$. The light probability matrices for $k\in\{k_1,k_2,k_3,k_4\}$ are very close at this level.}
\centering
  \begin{tabular}{lllll}
    \hline\hline
     & \textbf{d[1]} & \textbf{d[2]} & \textbf{d[3]} & \textbf{d[4]} \\\hline
    m=0.0 & 1.1e-6 & 0.00013 & 0.012 & 0.9 \\
    m=-0.083 & 1.4e-6 & 0.00014 & 0.014 & 0.83 \\\hline\hline
  \end{tabular}
\end{table}

\subsection{Results in 4D}

The results in 4D are qualitatively very similar to those in 3D, so we only offer some brief comments.

\subsubsection{N=16}

In 4D for $N=16$ we compute to higher values of $k$ than in 3D because signs of convergence are seen at level $L=8$ for higher values of $k$ up to $10000$. \Cref{tab:table-0D4} shows that when both the interval integration and the midpoint approximation methods are efficiently applicable, they yield close results at the $L=8$. The interval integration method becomes unfeasible at $k\sim 1000$, so in the following data tables we show results from the midpoint approximation method.

\begin{table}[H]
\caption{\label{tab:table-0D4}Interval integration vs. midpoint approximation methods for $D=4, N=16, L=8$. The table shows $d(p,p')$ for light probability matrices $p$ and $p'$ evaluated using the two methods. At $k=1000$ and above the interval integration method becomes inefficient so the comparison is not made.} 
\centering
  \begin{tabular}{llllllll}
    \hline\hline
     & \textbf{k=0.0} & \textbf{k=1.0} & \textbf{k=1e1} & \textbf{k=1e2} & \textbf{k=1e3} & \textbf{k=1e4} & \textbf{k=1e5} \\\hline
    m=0.0 & 0.0 & 1.3e-11 & 1.3e-9 & 1.3e-7 & N/A & N/A & N/A \\
    m=-0.083 & 0.097 & 0.097 & 0.097 & 0.097 & N/A & N/A & N/A \\\hline\hline
  \end{tabular}
\end{table}

\begin{table}[H]
\caption{\label{tab:table-1D4}Convergence measured by $d(p_L,p_{L-1})$   as the refinement level $L$ increases for $D=4, N=16$.}
\centering
  \begin{tabular}{llllllll}
    \hline\hline
    \textbf{Level L} & \textbf{2} & \textbf{3} & \textbf{4} & \textbf{5} & \textbf{6} & \textbf{7} & \textbf{8} \\\hline
    m=0.0, k=0.0 & 0.32 & 0.26 & 0.24 & 0.094 & 0.056 & 0.033 & 0.019 \\
    m=0.0, k=1.0 & 0.32 & 0.26 & 0.24 & 0.094 & 0.056 & 0.033 & 0.019 \\
    m=0.0, k=10.0 & 0.32 & 0.27 & 0.24 & 0.094 & 0.056 & 0.033 & 0.019 \\
    m=0.0, k=100.0 & 0.35 & 0.57 & 0.27 & 0.097 & 0.056 & 0.033 & 0.019 \\
    m=0.0, k=1000.0 & 0.58 & 0.63 & 0.48 & 0.23 & 0.074 & 0.033 & 0.019 \\
    m=0.0, k=10000.0 & 0.55 & 0.41 & 0.87 & 0.4 & 0.97 & 0.59 & 0.024 \\
    m=0.0, k=100000.0 & 0.29 & 0.59 & 0.55 & 0.42 & 0.32 & 0.81 & 0.88 \\
    m=-0.083, k=0.0 & 0.33 & 0.25 & 0.25 & 0.15 & 0.084 & 0.056 & 0.034 \\
    m=-0.083, k=1.0 & 0.33 & 0.25 & 0.25 & 0.15 & 0.084 & 0.056 & 0.034 \\
    m=-0.083, k=10.0 & 0.32 & 0.26 & 0.25 & 0.15 & 0.084 & 0.056 & 0.034 \\
    m=-0.083, k=100.0 & 0.31 & 0.54 & 0.22 & 0.15 & 0.084 & 0.056 & 0.034 \\
    m=-0.083, k=1000.0 & 0.72 & 0.71 & 0.3 & 0.32 & 0.1 & 0.056 & 0.034 \\
    m=-0.083, k=10000.0 & 0.71 & 0.44 & 0.7 & 0.29 & 0.77 & 0.46 & 0.042 \\
    m=-0.083, k=100000.0 & 0.33 & 0.54 & 0.48 & 0.27 & 0.49 & 0.5 & 0.83 \\\hline\hline
  \end{tabular}
\end{table}

\begin{table}[H]
\caption{\label{tab:table-2D4}For $D=4, N=16, L=8$ and $k$ values $(k_1,k_2,k_3,k_4,k_5,k_6,k_7):=(0,1,10,100,1000,10000,100000)$, showing $d[i]:=d(p_{k_i},p_{k_{i+1}})$. The light probability matrices for $k\in\{k_1,k_2,k_3,k_4,k_5,k_6\}$ are very close at this level.}
\centering
  \begin{tabular}{lllllll}
    \hline\hline
     & \textbf{d[1]} & \textbf{d[2]} & \textbf{d[3]} & \textbf{d[4]} & \textbf{d[5]} & \textbf{d[6]} \\\hline
    m=0.0 & 8.5e-12 & 8.4e-10 & 8.4e-8 & 1.2e-5 & 0.00087 & 0.47 \\
    m=-0.083 & 1.1e-11 & 1.1e-9 & 1.1e-7 & 1.1e-5 & 0.0012 & 0.31 \\\hline\hline
  \end{tabular}
\end{table}

\begin{figure}[H]
    \centering
    \includegraphics[width=1.\textwidth]{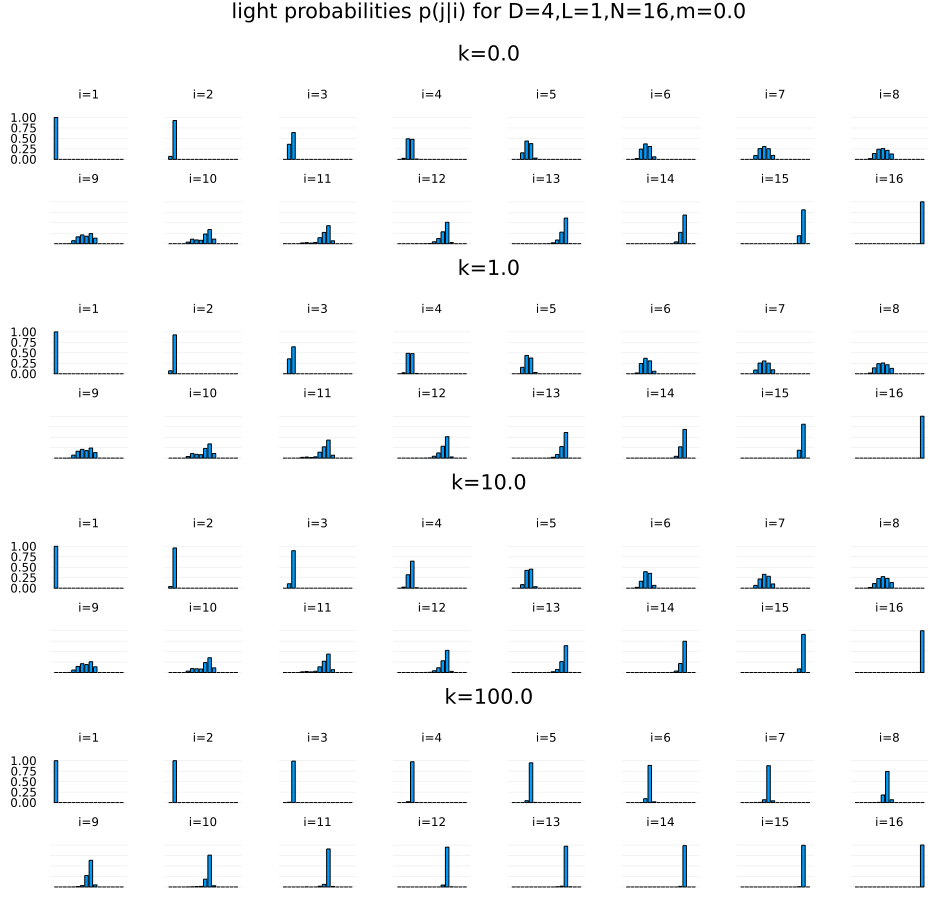}
    \caption{Initial light probability matrices without refinement as evaluated in the interval integration method. As $k$ is increased the probabilities are more sharply peaked.}
    \label{fig:probMat_D4-L1-N16-m0}
\end{figure}

\begin{figure}[H]
    \centering
    \includegraphics[width=1.\textwidth]{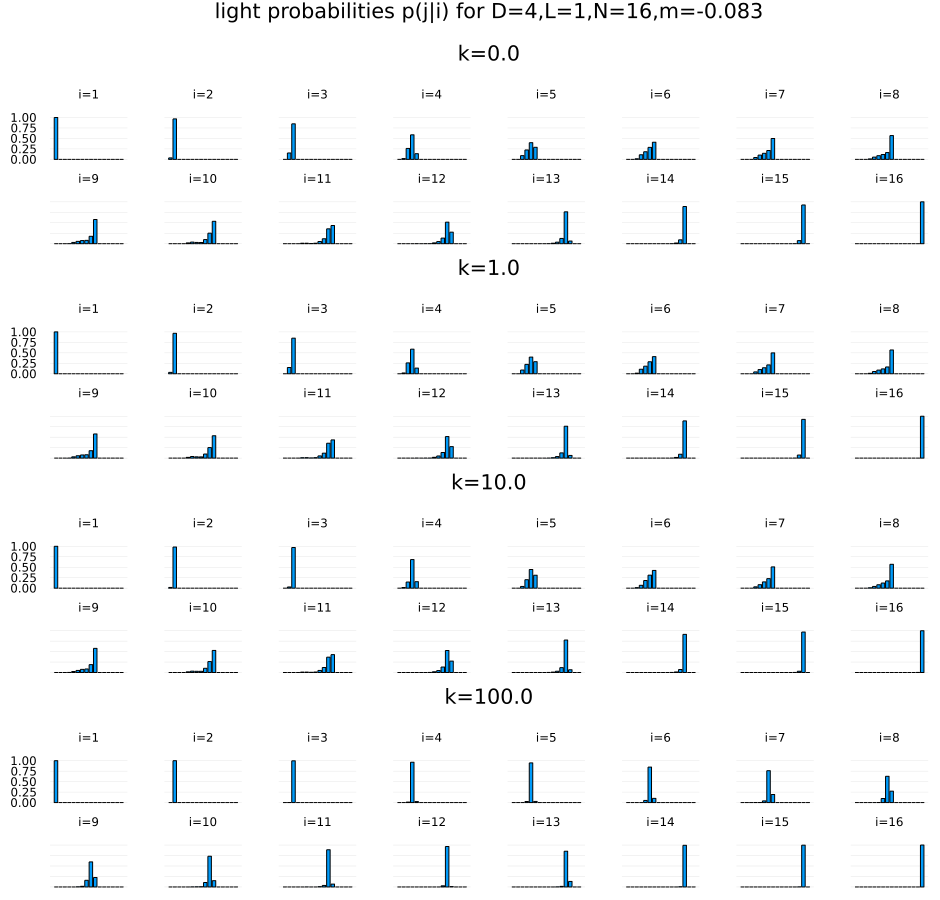}
    \caption{Initial light probability matrices without refinement as evaluated in the interval integration method. As $k$ is increased the probabilities are more sharply peaked.}
    \label{fig:probMat_D4-L1-N16-m-}
\end{figure}

\begin{figure}[H]
    \centering
    \includegraphics[width=1.\textwidth]{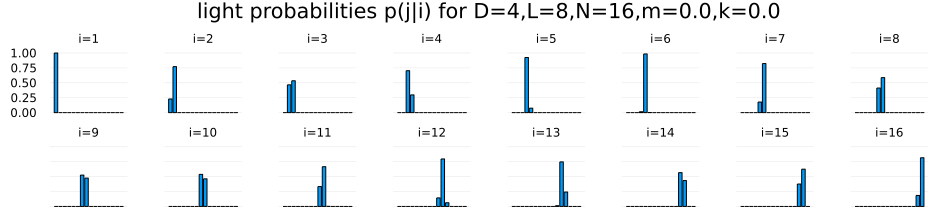}
    \caption{The commonly approached light probability matrix for $N=16,m=0$.}
    \label{fig:probMat_D4-L8-N16-m0}
\end{figure}

\begin{figure}[H]
    \centering
    \includegraphics[width=1.\textwidth]{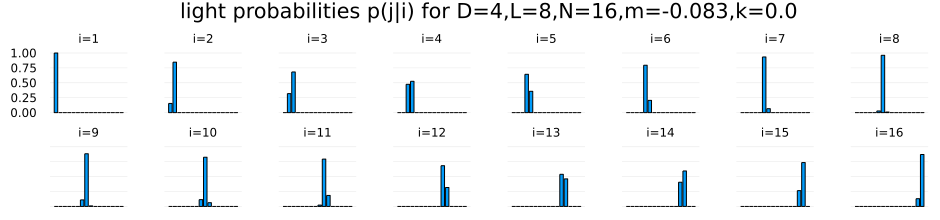}
    \caption{The commonly approached light probability matrix for $N=16,m=-1/12$.}
    \label{fig:probMat_D4-L8-N16-m-}
\end{figure}

Without refinement, the light probability matrices exhibit different shapes (\Cref{fig:probMat_D4-L1-N16-m0} for $m=0$; \Cref{fig:probMat_D4-L1-N16-m-} for $m=-1/12$). As the lattice is refined, signs of convergence of the light probability matrices are seen across $k=0.0$ to $k\sim 10000$ (\Cref{tab:table-1D4}, \Cref{tab:table-2D4}, \Cref{fig:probMat_D4-L8-N16-m0}, \Cref{fig:probMat_D4-L8-N16-m-}). Signs of convergence at the refinement level $L=8$ are not seen once $k$ reaches $100000$ (\Cref{tab:table-1D4}, \Cref{tab:table-2D4}).

As in 3D, a comparison between light probabilities at the no refinement level (\Cref{fig:probMat_D4-L1-N16-m0}, \Cref{fig:probMat_D4-L1-N16-m-}) and those at the highest refinement level (\Cref{fig:probMat_D4-L8-N16-m0}, \Cref{fig:probMat_D4-L8-N16-m-}) shows that that refinement drives both very wide and very narrow distributions to an intermediate one.

\subsubsection{N=32}

\begin{figure}[H]
    \centering
    \includegraphics[width=1.\textwidth]{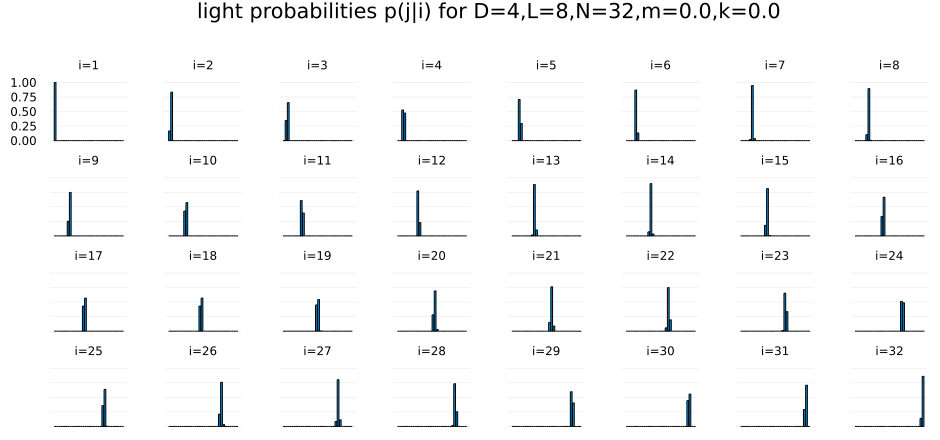}
    \caption{The commonly approached light probability matrix for $N=32,m=0$.}
    \label{fig:probMat_D4-L8-N32-m0}
\end{figure}

\begin{figure}[H]
    \centering
    \includegraphics[width=1.\textwidth]{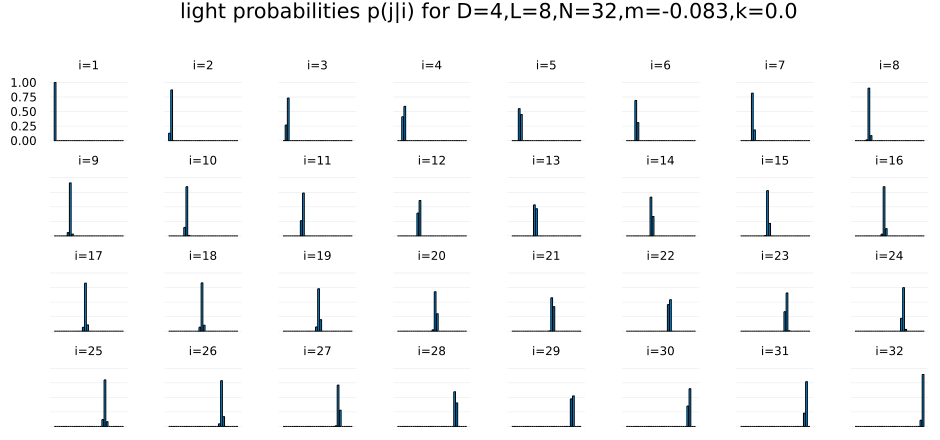}
    \caption{The commonly approached light probability matrix for $N=32,m=-1/12$.}
    \label{fig:probMat_D4-L8-N32-m-}
\end{figure}

For $N=32$ the numerical results are shown in \Cref{tab:table-0D4N32}, \Cref{tab:table-1D4N32}, \Cref{tab:table-2D4N32}, \Cref{fig:probMat_D4-L8-N32-m0}, and \Cref{fig:probMat_D4-L8-N32-m-}. Different from the $N=16$ case, at level $L=8$ convergence of the light probability matrices is not seen at $k_4=1000$ and above (In \Cref{tab:table-1D4N32} the two rows of $k_6=10000$ showed small $d(p_L,p_{L-1})$ at $L=8$. However, at $L=7$ the same quantity is still large, and it is unclear from the available data if convergence has taken place at $k_6=10000$.). 

The reason for the difference between the $N=32$ and the $N=16$ cases is unclear. It could well be that as $N$ increases, a higher level of refinement with finer resolution is needed to see the convergence. On the other hand, signs of convergence are observed in the range $k=0$ to $k\sim 100$ for $N=32$, which suggest universal behavior also in this case.

\begin{table}[H]
\caption{\label{tab:table-0D4N32}Interval integration vs. midpoint approximation methods for $D=4, N=32, L=8$. The table shows $d(p,p')$ for light probability matrices $p$ and $p'$ evaluated using the two methods. At $k=1000$ and above the interval integration method becomes inefficient so the comparison is not made.} 
\centering
  \begin{tabular}{llllllll}
    \hline\hline
     & \textbf{k=0.0} & \textbf{k=1.0} & \textbf{k=1e1} & \textbf{k=1e2} & \textbf{k=1e3} & \textbf{k=1e4} & \textbf{k=1e5} \\\hline
    m=0.0 & 0.0 & 1.3e-11 & 1.3e-9 & 1.3e-7 & N/A & N/A & N/A \\
    m=-0.083 & 0.082 & 0.082 & 0.082 & 0.082 & N/A & N/A & N/A \\\hline\hline
  \end{tabular}
\end{table}

\begin{table}[H]
\caption{\label{tab:table-1D4N32}Convergence measured by $d(p_L,p_{L-1})$ as the refinement level $L$ increases for $D=4, N=32$.} 
\centering
  \begin{tabular}{llllllll}
    \hline\hline
    \textbf{Level L} & \textbf{2} & \textbf{3} & \textbf{4} & \textbf{5} & \textbf{6} & \textbf{7} & \textbf{8} \\\hline
    m=0.0, k=0.0 & 0.21 & 0.3 & 0.23 & 0.23 & 0.13 & 0.1 & 0.067 \\
    m=0.0, k=1.0 & 0.21 & 0.3 & 0.23 & 0.23 & 0.13 & 0.1 & 0.067 \\
    m=0.0, k=10.0 & 0.18 & 0.3 & 0.26 & 0.23 & 0.13 & 0.1 & 0.067 \\
    m=0.0, k=100.0 & 0.46 & 0.56 & 0.71 & 0.85 & 0.66 & 0.37 & 0.071 \\
    m=0.0, k=1000.0 & 0.61 & 0.64 & 0.8 & 0.93 & 0.72 & 0.75 & 1.0 \\
    m=0.0, k=10000.0 & 0.65 & 0.75 & 0.34 & 0.6 & 0.98 & 0.61 & 0.075 \\
    m=0.0, k=100000.0 & 0.69 & 0.89 & 0.73 & 0.85 & 0.93 & 0.99 & 1.0 \\
    m=-0.083, k=0.0 & 0.34 & 0.39 & 0.28 & 0.23 & 0.16 & 0.11 & 0.081 \\
    m=-0.083, k=1.0 & 0.34 & 0.39 & 0.28 & 0.23 & 0.16 & 0.11 & 0.081 \\
    m=-0.083, k=10.0 & 0.25 & 0.36 & 0.26 & 0.24 & 0.16 & 0.11 & 0.081 \\
    m=-0.083, k=100.0 & 0.45 & 0.68 & 0.68 & 0.77 & 0.77 & 0.34 & 0.087 \\
    m=-0.083, k=1000.0 & 0.55 & 0.61 & 0.83 & 0.93 & 0.81 & 0.43 & 0.97 \\
    m=-0.083, k=10000.0 & 0.52 & 0.6 & 0.58 & 0.62 & 0.81 & 0.7 & 0.09 \\
    m=-0.083, k=100000.0 & 0.46 & 0.72 & 0.53 & 0.84 & 0.77 & 0.81 & 1.0 \\\hline\hline
  \end{tabular}
\end{table}

\begin{table}[H]
\caption{\label{tab:table-2D4N32}For $D=4, N=32, L=8$ and $k$ values $(k_1,k_2,k_3,k_4,k_5,k_6,k_7):=(0,1,10,100,1000,10000,100000)$, showing $d[i]:=d(p_{k_i},p_{k_{i+1}})$. The light probability matrices for $k\in\{k_1,k_2,k_3,k_4\}$ are very close at this level.}
\centering
  \begin{tabular}{lllllll}
    \hline\hline
     & \textbf{d[1]} & \textbf{d[2]} & \textbf{d[3]} & \textbf{d[4]} & \textbf{d[5]} & \textbf{d[6]} \\\hline
    m=0.0 & 1.1e-6 & 0.00013 & 0.012 & 0.9 & 0.66 & 0.67 \\
    m=-0.083 & 1.4e-6 & 0.00014 & 0.014 & 0.83 & 0.88 & 0.54 \\\hline\hline
  \end{tabular}
\end{table}

\section{Discussions}\label{sec:d}

Lorentzian gravitational path integrals offer both challenges and opportunities. On the one hand, they are harder to compute because the complex amplitudes leads to the numerical sign problem. On the other hand, they supply physical quantities associated with causal structures which are not available in Euclidean theories.

% In the previous work of \cite{Jia2022LightGravity} light ray fluctuation is studied in symmetry-reduced box models with just one dynamical degree of freedom. The current results are an improvement that incorporates many dynamical degree of freedom.

In this work we took advantage of one such quantity, the light ray probabilities, to study the lattice refinement limit of Lorentzian simplicial quantum gravity. To overcome the computational challenge of the Lorentzian theory, the study is based in a simplified refined box model setting. This is a lattice containing many dynamical edges (\Cref{fig:refine_box}), but with the simplifications of:
\begin{enumerate}
    \item Frozen frame. All the interior horizontal and vertical edge squared lengths are kept fixed.
    \item Partial symmetry. Within an elementary box the diagonal interior edges share the same squared length.
\end{enumerate}
For a range of the gravitational coupling constant values, the  numerical results for the light ray probabilities show signs of convergence upon lattice refinement. Moreover, the light ray probabilities from this range of gravitational coupling values actually approach a common distribution upon lattice refinement. This suggests the presence of a universality class.

How should we interpret these results for the study of the lattice refinement limit of Lorentzian simplicial quantum gravity? This section is a discussion on what has been achieved, and what still lies ahead.

\subsection{Microscopically large and macroscopically small light ray fluctuations}

Intuitively, quantum spacetime fluctuations are expected to be large microscopically and small macroscopically. The above results meet this intuition, since microscopically large light ray fluctuations in the elementary boxes give rise to macroscopically small light ray fluctuations. 

To see this, we note a result from  \cite{Jia2022LightGravity} that the light probabilities for the elementary boxes obey
\begin{align}\label{eq:sr3D}
p[l^2\sigma_B,k] =& p[\sigma_B,l k], \quad \text{in 3D},\\
p[l^2\sigma_B,k] =& p[\sigma_B,l^2 k], \quad \text{in 4D}.\label{eq:sr4D}
\end{align}
These say that scaling the boundary squared lengths $\sigma_B$ by a factor of $l^2$ amounts to scaling the coupling $k$ by $l$ in 3D or $l^2$ in 4D. \Cref{fig:probMat_D3-L1-N16-m0} and \Cref{fig:probMat_D4-L1-N16-m0} show that light fluctuations are significant for small values of $k$, as one should expect since $k$ contains a factor of $1/\hbar$ which becomes small in the quantum limit $\hbar>>1$ where we expect large quantum fluctuations. Although the probabilities of \Cref{fig:probMat_D3-L1-N16-m0} and \Cref{fig:probMat_D4-L1-N16-m0} are for boxes of the default boundary size, from \eqref{eq:sr3D} and \eqref{eq:sr4D} we see that decreasing the boundary size by $l<1$ amounts to keeping the boundary size but decreasing $k$ by a factor of $l$ in 3D and $l^2$ in 4D. For any finite $k$, some sufficiently large lattice refinement level $L$ will assign a sufficiently small $l$ so that the elementary boxes exhibit significant light ray fluctuations. From results such as in \Cref{fig:probMat_D3-L8-N16-m0}, \Cref{fig:probMat_D3-L8-N16-m-}, \Cref{fig:probMat_D4-L8-N16-m0}, and \Cref{fig:probMat_D4-L8-N16-m-}, we see that composing the elementary boxes produces light ray probabilities with smaller fluctuations in the macroscopic.

That microscopically large light ray fluctuations in the elementary boxes give rise to macroscopically small light ray fluctuations in the refinement limit is quite encouraging. It demonstrates that the converging numerical results exhibit reasonable physical properties. 

\subsection{Pinpointing the universality class}\label{sec:oub}

We just noted that for any $k<\infty$, there is some lattice refined enough to exhibit large light ray fluctuations in the elementary boxes. This turns out to help us understand the putative universality class noted in \Cref{sec:uc} better.

Looking at \Cref{fig:pmflow} again, we recognise three regimes along the parameter space of $k\in [0,\infty]$. For $k=0$ up to $k\sim 100$, the light probabilities converge toward a common value upon lattice refinement. %This qualitative feature is seen in all the models studied, for spacetime dimensions $D=3,4$, for the number of intervals $N=16,32$, and for measure factor exponents $m=0,-1/12$.
The infinite $k=\infty$ limit corresponds to the classical $\hbar=0$ limit. We know without computation that there is no light ray fluctuation since the path integral is completely dominated by the stationary point, which is the flat spacetime configuration in the present setting. Lattice refinement does not change this infinity-enhanced domination, so $k=\infty$ is a refinement fixed point. 

The third regime is between $k\sim 100$ and $k=\infty$. We have no numerical evidence whether there are refinement fixed points here or not. Yet here is an analytic reasoning suggesting that the whole range of $k\in[0,\infty)$ belongs to the same putative universality class, so actually there is no third regime.
%In principle, $k^*$ could go all the way up to $\infty$ so that the universality class encompasses $k\in[0,\infty)$ (like in 1D Ising model with the Hamiltonian $H=-K\sum_{<i,j>}s_i s_j$, the zero temperature limit $K=\infty$ constitutes an unstable renormalization group coarsening fixed point, while $K\in[0,\infty)$ all flow to $K=0$ in one universality class \cite{Cardy1996ScalingPhysics}). Or $k^*$ could be finite (like in 2D Ising model, the universality class $K\in[0,K^*)$ stops at some finite $K^*$  \cite{Cardy1996ScalingPhysics}).
We noted above that for any $k<\infty$, a high-level refinement can always make the elementary boxes exhibit large light ray fluctuations equivalent to a standard-sized box with as small a $k$ as possible. If indeed the universality class exists so that models with small $k$ all flow to a common fixed point, then composing these elementary boxes coming from any large finite $k$ will lead to the same fixed point. This would show that all $k\in[0,\infty)$ belong to the same universality class.

This reasoning almost offers a proof that the whole range of $k\in[0,\infty)$ belongs to the same putative universality class, but the assumption that models with small $k$ belong to the same universality class is so far only supported numerically. Perhaps with some further work one could find an analytic proof (or disproof) for the latter point.

\subsection{Limitations of the frozen frame}\label{sec:liff}

The above discussion on the boundary of the putative universality class in $k$ space suggests that light ray fluctuations in the present setting are scale invariant. If the universality class exists at all, then in the lattice refinement limit the light ray probabilities are the same for all finite values of $k$.

This highlights a limitation of the refined box model with the frozen frame assumption. In \Cref{sec:rb} we mentioned that the frozen frame assumption is suitable when the flat configuration dominates the path integral. In the present setting with just the Einstein-Hilbert term in the action and with the flat boundary condition, certainly the flat configuration should dominate the path integral. However, in the full theory we expect that $k$ sets a length scale $s^*$ so that fluctuations around the flat geometry below that scale make comparable contributions to the path integral. The refined box model allows fluctuations with the dynamical diagonal edges, but ignores fluctuations at the horizontal and vertical edges. That the frame is frozen for arbitrarily small elementary boxes seems to be the limitation that led to scale invariance, which prevented the scale $s^*$ to show up in the light ray probabilities in the refinement limit.

\subsection{Outlooks}

In summary, the present results are encouraging in the numerical evidences they provide for the lattice refinement fixed points and in the reasonable behavior for associated light ray fluctuations, reflecting large quantum spacetime fluctuations in the microscopic, and small quantum spacetime fluctuations in the macroscopic. On the other hand, the simplifying assumptions of the refined box model still need to be relaxed to mimic the full theory.

We speculate that once the frozen frame and the partial symmetry assumptions are relaxed, models with different $k$ values should exhibit at least three broad types of light ray probabilities in the lattice refinement limit. First, $k=\infty$ (and perhaps values close to it) should still exhibit no light ray fluctuation, because still the flat configuration overwhelmingly dominates over all other configurations. Second, $k=0$ and values close to it should exhibit wild light ray fluctuations in the lattice refinement limit than in \Cref{sec:uc}. This is because a small $k$ is not enough to maintain flat configuration dominance without the frozen frame assumption. Third, $k$ values between the above cases are expected to exhibit similar light ray fluctuations in the lattice refinement limit as in \Cref{sec:uc}. Without the frozen frame assumption, we expect that at a fine scale there is much light ray fluctuations, like in the $k\approx 0$ cases of \Cref{sec:uc}. In principle this can be checked by computing the light ray probabilities in the full theory for a box with very small boundary lengths. The results of \Cref{sec:uc} suggest that as we move to larger scales set by the boundary, the light ray fluctuations would become smaller. While the first two regimes have too little and too much fluctuations, this putative third regime has a chance to be realistic.

How to check if these speculations are realized when the frozen frame and the partial symmetry assumptions are relaxed? The challenge is, as always, the difficulty in evaluating multi-dimensional integrals with complex integrands. One possible way to proceed is through numerical computations that overcome the sign problem \cite{Jia2022ComplexProspects, ItoTensorCalculus}. Alternatively, the results of this paper show that the $k=0$ models approach the same light probability matrix upon refinement as the $k>0$ models. The $k=0$ case could help the computation, because here the integrand is real and non-negative.

As a first step, one could relax the partial symmetry assumption while maintaining the frozen frame assumption. In the $k=0$ case Monte Carlo simulations are efficient, and one could compute the light probabilities to see if there is a parameter range of $m$ where the lattice refinement leads to converging results. The result could belong to a universality class incorporating $k\ne 0$ cases.

\begin{figure}%[h]
    \centering
    \includegraphics[width=.8\textwidth]{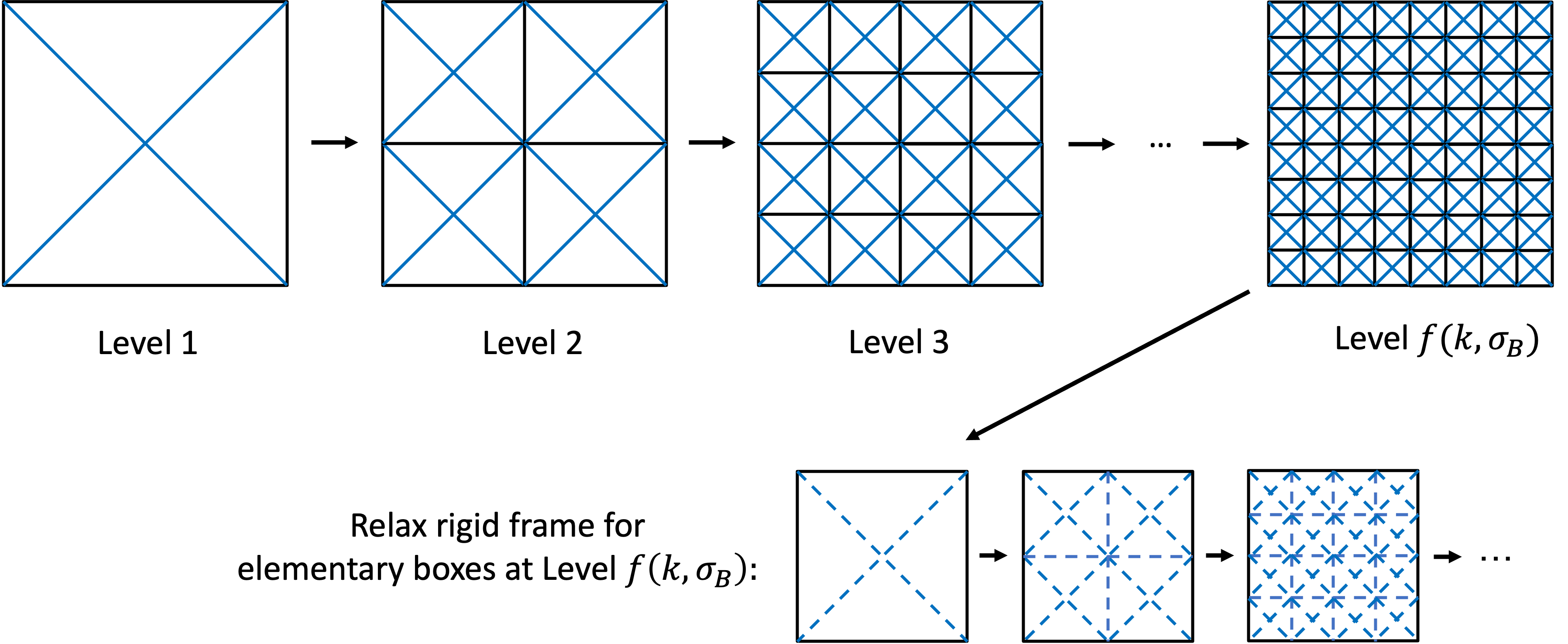}
    \caption{Lattice refinement limit.}
    \label{fig:scale_refine}
\end{figure}

%As illustrated in \Cref{fig:scale_refine}, one possibility to incorporate the length scale $s^*$ is to introduce a ``cutoff'' refinement level $f(k,\sigma_B)$ set by the coupling constant $k$ and the boundary squared lengths $\sigma_B$. 
Moreover, \Cref{fig:scale_refine} illustrates a scenario to incorporate the length scale $s^*$ discussed in \Cref{sec:liff}. Here we impose a ``cutoff'' refinement level $f(k,\sigma_B)$ set by the coupling constant $k$ and the boundary squared lengths $\sigma_B$. Beyond this level, further refinements of the elementary boxes allow wild small scale fluctuations by making dynamical the horizontal and vertical frame edges (second line in \Cref{fig:scale_refine}). If a refinement fixed point exists for the elementary boxes, then the composition of the elementary boxes at level $f(k,\sigma_B)$ would yield the wanted light probabilities for the total box region. Since different $k$ values lead to different level $f(k,\sigma_B)$ and hence different numbers of elementary boxes to compose, the light probabilities for the total box region will exhibit a dependence on $k$. This makes the scenario non scale-invariant.

How to check if a refinement fixed point exists for the elementary boxes? Note that the boundary lengths for the elementary boxes at the wild fluctuation scale $s^*$ are expected to be tiny in realistic cases. By \eqref{eq:sr3D} and \eqref{eq:sr4D}, after scaling the boundaries to the standard box size, the effective $k$ value ($lk$ in 3D and $l^2k$ in 4D) is expected to be tiny. The results of this paper suggest that if $k^*$ is sufficiently tiny, the light probabilities for this elementary box belongs to the same universality class of the $k=0$ model. Whether a refinement fixed point exist in the latter can be checked efficiently by Monte Carlo simulation since the $k=0$ model has a non-negative integrand. 

% Furthermore, one could compute the light probabilities by Monte Carlo simulations in the $k=0$ case with both the partial symmetry and the frozen frame assumptions relaxed. Refining the lattice may yield a converging result exhibiting wildly fluctuating light probabilities. We could interpret this as pertaining to a box region of a $k>0$ model at a fine scale $s^*$ set by $k$, as mentioned in \Cref{sec:liff}. Composing many such boxes may mimic the light probabilities for the $k>0$ model at a larger scale. The larger the boundary lengths for this large scale, the more boxes to be composed. The larger $k$ is, the more boxes to be composed (recall \eqref{eq:sr3D} and \eqref{eq:sr4D}). This way, one could possibly obtain scale-dependent light probabilities, in contrast to the present scale-invariant results. In this setting, the continuum limit is hoped to be reached by finding converging light probabilities for the $k=0$ case under lattice refinement.

These set a concrete path to proceed for future works. Of course, these models are still specialized to include only the Einstein-Hilbert term in the action, and come without matter coupling. Further studies could relax these assumptions. 

Last but certainly not least, let us not forget that quantum gravitational light ray fluctuations are not just computable, but also potentially measurable. To realize this potential, we should probably look at highly non-perturbative effects such as quantum gravitational tunnelling around would-be singularities \cite{JiaIsTrivial}. It is worth investigating light ray fluctuations for the boundary conditions of these events.

\section*{Acknowledgement}

I am very grateful to Lucien Hardy and Achim Kempf for long-term encouragement and support. Research at Perimeter Institute is supported in part by the Government of Canada through the Department of Innovation, Science and Economic Development Canada and by the Province of Ontario through the Ministry of Economic Development, Job Creation and Trade. 

% \appendix

\bibliographystyle{unsrt}
\bibliography{mendeley.bib}

\end{document}